\documentclass[aps,prb,floatfix,twocolumn,showpacs,reprint,superscriptaddress]{revtex4-1}
\usepackage{float}
\usepackage{graphicx}
\usepackage{amsmath}
\usepackage{bbold}

\DeclareMathOperator{\Tr}{Tr}

\DeclareMathOperator{\sgn}{sgn}

\begin{document}
\title{Critical Behavior of Four-Terminal Conductance of Bilayer Graphene Domain Walls}

\author{Benjamin J. Wieder}
\affiliation{Department of Physics and Astronomy, University of Pennsylvania,
Philadelphia, PA 19104}

\author{Fan Zhang}
\affiliation{Department of Physics, University of Texas at Dallas, Richardson, TX 75080}

\author{C. L. Kane}
\affiliation{Department of Physics and Astronomy, University of Pennsylvania,
Philadelphia, PA 19104}

\begin{abstract}
Bilayer graphene in a perpendicular electric field can host domain walls between regions of reversed field direction or interlayer stacking. The gapless modes propagating along these domain walls, while not strictly topological, nevertheless have interesting physical properties, including valley-momentum locking. A junction where two domain walls intersect forms the analogue of a quantum point contact. We study theoretically the critical behavior of this junction near the pinch-off transition, which is controlled by two separate classes of non-trivial quantum critical points.  For strong interactions, the junction can host phases of unique charge and valley conductances.  For weaker interactions, the low-temperature charge conductance can undergo one of two possible quantum phase transitions, each characterized by a specific critical exponent and a collapse to a universal scaling function, which we compute.

\end{abstract} 

\pacs{74.45.+c, 71.10.Pm, 74.78.Fk, 74.78.Na}
\maketitle

\section{Introduction}
\label{sec:introduction}

\subsection{Background}
\label{sec:back}

Bilayer graphene~\cite{review,McCann} provides an attractive platform for unconventional two-dimensional electronic physics
due to the two quadratic band contacts at its Fermi points,
and because of the variety of ways by which one can introduce a band gap and produce momentum-space Berry curvature~\cite{SQH-Zhang}.  The interlayer nearest-neighbor hopping, $\gamma_1$, warps the band structure of the individual graphene layers, repelling two bands away from the Fermi energy and leaving the remaining two dispersing quadratically.  This warping is a consequence of the two-step process in which electrons hop between the two low-energy sublattices via the two high-energy ones.  The band touching points at inequivalent Brillouin zone corners $K$ or $K'$ are protected by the Berry phase $\pm 2\pi$ (or winding number $\pm 2$) and the required chiral (or sublattice) symmetry between the low-energy sublattices on opposite layers~\cite{Velasco}.  Keeping all of these symmetries, the band touching point can only split, instead of being gapped out, even when trigonal warping and other weak remote-hopping processes are taken into account~\cite{McCann}.

However, the chiral symmetry between the low-energy sublattices can be intrinsically broken by spin-orbit coupling~\cite{SOC}, spontaneously broken by electron-electron interactions~\cite{SQH-Zhang,RG-Zhang,MF-Levitov}, or explicitly broken by an interlayer potential difference~\cite{McCann,Ohta,Oostinga,Min}.  As a consequence, the quadratic band touching is no longer symmetry-protected and gaps open up at valleys $K$ and $K'$. While the first two types of gaps are small in practice~\cite{Velasco,Bao,Min1,Yugui}, the electric-field-induced gap, which is the focus of this paper, saturates at a large value comparable to the interlayer hopping $\gamma_1\sim 0.3$~eV~\cite{Min,Yuanbo}.  Opening the band gap produces large momentum-space Berry curvature in the quasiparticle bands, with the curvature integral quantized to $\pm 1$ over a half Brillouin zone centered at $K$ or $K'$~\cite{SQH-Zhang,Martin}.
Moreover, for bilayer graphene gapped by an electric field, the sign of this partial Chern number depends on the valley index, the sign of the energy gap (given by the direction of interlayer electric field), and the layer stacking order (i.e., AB or BA)~\cite{SQH-Zhang,LSW}.  Here AB (BA) stacking refers to the case in which $\gamma_1$ couples the top layer A (B) and bottom layer B (A) high-energy sublattices.

In the presence of an interlayer electric field, when the field direction is reversed across a line~\cite{Martin,LSW,Yao,Jung,Qiao,Peeters,Li}
or when the field is uniform and the layer stacking switches from AB to BA~\cite{LSW,Alden,Vaezi}, the valley-projected Chern number changes by $2$ ($-2$) across the domain wall in valley $K$ ($K'$).  As a result, both types of domain walls host two chiral edge states in each valley with chirality (direction) locked to valley index $K$ or $K'$, as shown in Fig.~\ref{FigWall}. Similar domain wall states also occur spontaneously due to interactions in the absence of electric fields but at finite temperature~\cite{XiaoLi}.  Importantly~\cite{LSW}, these ``Quantum Valley Hall'' (QVH) edge states are not strictly topological and can be gapped out by a sufficiently strong, large-momentum scattering which couples the two valleys, even if the underlying symmetries are still preserved.  It is therefore crucial that valley index also remains a ``good quantum number,'' for which we will assume that short-range disorder, interlayer stacking, and electric field direction changes are smooth on the scale of the lattice.  Under this assumption, backscattering is prohibited and the system of domain walls provides an attractive platform for Tomonaga-Luttinger liquid physics~\cite{affleck}. 

In this paper, we study the electronic transport properties of a junction where two domain walls intersect (Fig.~\ref{FigQPC}).  Such a structure resembles the quantum point contact of the edges of two Quantum Spin Hall (QSH) insulators, which has been studied in Refs.~\onlinecite{houkim,strom,teokane} and can be probed by four-terminal transport measurements.  A domain wall junction can be tuned through a ``pinch-off transition'' by applying a local field (such as a perpendicular electric field) to the junction region.  In Ref.~\onlinecite{teokane}, it was found that the corresponding pinch-off transition for QSH systems is controlled by a novel quantum critical point, and that at low temperatures the conductance is described by a universal scaling function across the pinch-off transition.  In contrast to the QSH edge states, which have a single time-reversed pair of helical modes, the domain wall states in bilayer graphene host four helical pairs (including electron spin).  We find that this leads to several important modifications of the low-energy properties of the junction.  

Unlike with the QSH edge states, whose forward-scattering interactions are characterized by a single Luttinger parameter, it has been argued that for the domain wall states in bilayer graphene, one should characterize interactions with two independent Luttinger parameters~\cite{affleck}.  This leads to an expanded phase diagram for the possible stable states of the junction.  Moreover, we find that the pinch-off transition is modified.  Depending on the interaction strengths, there are two possible regimes for the reduced, two-terminal conductance: one in which it undergoes a single pinch-off transition directly from $0$ to $8e^{2}/h$ and one in which it undergoes two separate transitions, separated by a stable state with conductance $4e^{2}/h$.  We study the critical behavior of these transitions and compute the universal crossover scaling functions for weak interactions.   

This paper is structured as follows.  First we introduce in detail the domain wall states in bilayer graphene and derive low-energy effective field theories for them.  Adding interactions, we show how these states are Luttinger liquids described by two independent Luttinger parameters.  From there, we characterize the geometry of two intersecting domain wall states in the language of the resulting effective quantum point contact.  We then analyze the resulting four-terminal junction in both the context of many-body tunneling in a bosonization framework and with an S-matrix renomalization group using diagrammatic perturbation theory.  Combining these analyses, we determine the behavior of the reduced, two-terminal conductance over a range of interaction strengths.

\subsection{Measurable Results}
\label{sec:resultsum}

In this paper, we calculate several measurable properties of bilayer graphene domain wall quantum point contacts.  In section~\ref{sec:domainwalls}, we find the critical exponent $\alpha_{T}$ which characterizes the low temperature tunneling conductance scaling for a single domain wall, a result previously derived in Ref.~\onlinecite{affleck}.  In~\ref{sec:fourterminal}, we introduce a diagonal conductance $G_{ZZ}=8e^{2}/h$, which is only strictly quantized when valley index is conserved both within individual domain walls and across the junction. This conductance, therefore, stands as a first test of whether experimental samples are in the appropriate disorder regime for the analysis in this paper.  We also show in~\ref{sec:fourterminal} that states of exotic charge and valley tunneling character dominate the conductance of the junction under very strong attractive or repulsive interactions (Fig.~\ref{FigPhase}).  Finally, in~\ref{sec:scaling}, we show that the left-to-right conductance $G_{XX}$ undergoes either a direct transition from $0$ to $8e^{2}/h$ or an indirect one with an intermediate step to $4e^{2}/h$ depending on experimental specifics (Fig.~\ref{FigZeroT}).  Labeling the direct transition $A$ and the first step of the indirect transition $B$, we go on to show that at low temperatures, the conductance transitions should collapse onto universal scaling functions $\mathcal{G}_{A/B}$ with critical exponents $\alpha_{A/B}$ as functions of the interaction strengths (Figs.~\ref{FigTemp},\ref{FigScaling}).

\section{Model System}
\label{sec:model}

In this section we introduce our model system of bilayer graphene domain wall modes.  First, we begin with the Hamiltonian for a single domain wall and the Luttinger liquid physics which govern it in the presence of interactions.  Then, we discuss the four-terminal geometry which arises at the intersection of two domain walls and its equivalence to a quantum point contact.

\subsection{Domain Walls in Bilayer Graphene}
\label{sec:domainwalls}

\begin{figure}
\centering
\includegraphics[width=3.5in]{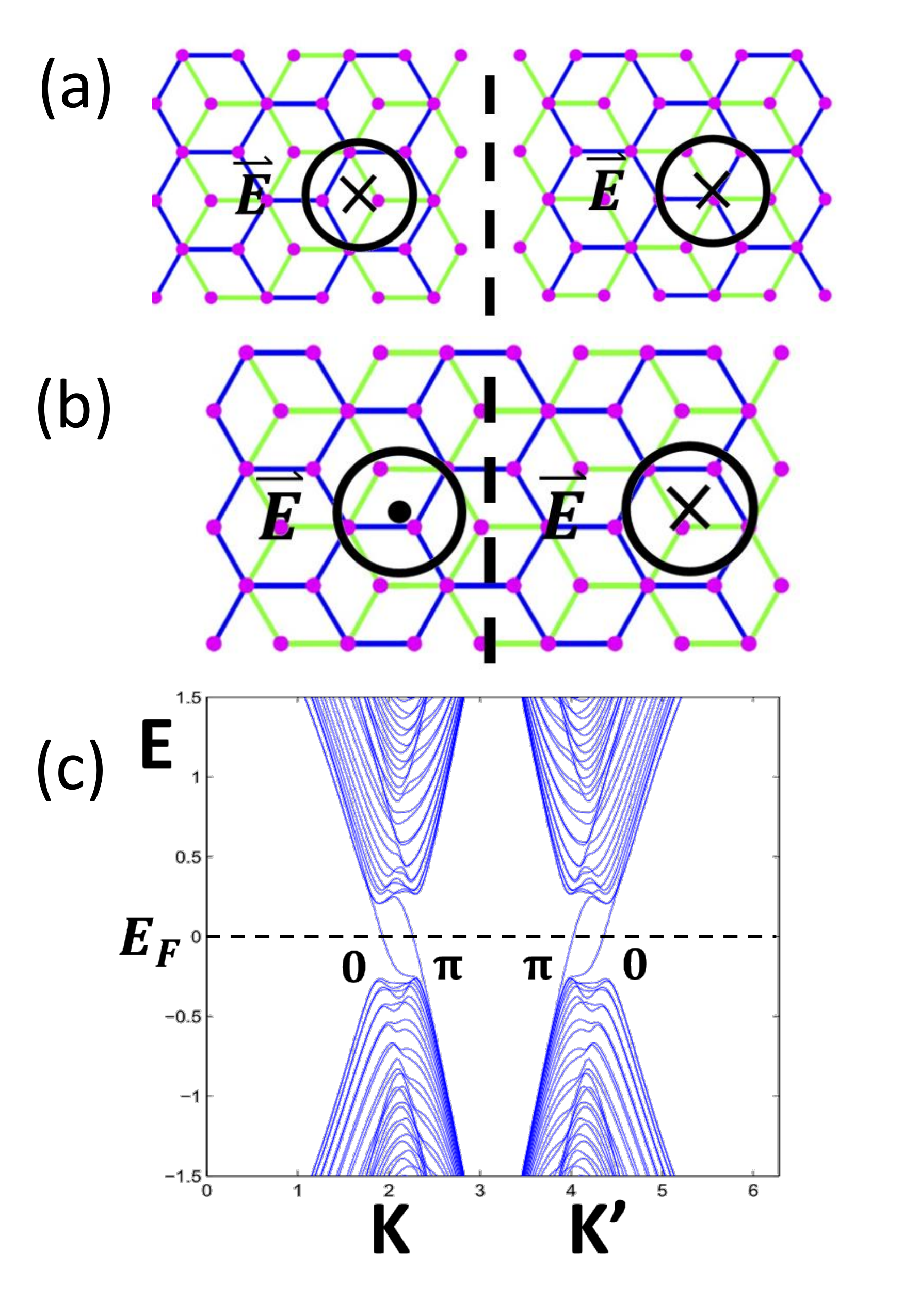}
\caption{Domain walls in bilayer graphene can be induced by applying a perpendicular electric field and varying either the interlayer stacking (a) or the electric field direction (b).  Both kinds of domain walls (the dotted lines) have similar domain wall band structures (c) when the Fermi energy $E_{F}$ is near the chiral symmetric point.  Adopting the notation of Ref.~\onlinecite{affleck}, the two domain wall states in each direction are labeled $0$ and $\pi$ working from the Brillouin zone edge in.  When the Fermi energy is exactly in the middle of the bulk gap, the Fermi velocities are the same for the $0$ and $\pi$ bands and electron direction is set by valley index $K$/$K'$.}
\label{FigWall}
\end{figure}

As discussed in the introduction, bilayer graphene domain walls can be created by varying the direction of the interlayer electric field or by varying the interlayer stacking order.  In either case, The valley-projected Chern number changes by $2$ ($-2$) across the domain wall in valley $K$ ($K'$).  This necessitates the existence of two domain wall states in each valley, with the states at $K$ having equal and opposite velocities to those at $K$'.  Adopting the notation of Ref.~\onlinecite{affleck}, we label these bands $0$ and $\pi$ respectively (Fig.~\ref{FigWall}).  For the purposes of our model, we will assume that the Fermi energy lies exactly in the middle of the bulk gap, which allows the simplification $v_{F,0} = v_{F,\pi}=v_{F}$.  This allows us to write down the non-interacting Hamiltonian density

\begin{equation}
\mathcal{H}_{0} = i\hbar v_{F}\sum_{a=0,\pi}\sum_{\sigma=\uparrow,\downarrow}\psi^{\dagger}_{a\sigma,in}\partial_{x}\psi_{a\sigma,in} - \psi^{\dagger}_{a\sigma,out}\partial_{x}\psi_{a\sigma,out}
\end{equation}

where the indexes ``in'' and ``out'' refer to direction and can correspond to electron operators with valley index $K$ or $K'$ depending on the orientation of the domain wall.  We will see later that in a four-terminal structure, the correspondence between in/out and $K/K'$ will alternate with lead index due to the valley-momentum locking of the domain wall modes.  

In the limit of the Fermi energy resting exactly at the chiral-symmetric point, the band indexes become arbitrary labels for all of the calculations in this paper.  A consequence of this is the emergence of a band-index-exchange symmetry in this problem, which we will frequently highlight in calculations throughout this paper.  Deviations from this point in the Fermi energy are expected in physical systems, and will lead in general to a relaxation of this symmetry.  If the deviations are small, the physics should resemble the predictions of this paper, with corrections of order $\sim 1-v_{F,0}/v_{F,\pi}$.  These corrections greatly complicate the calculations in this paper and obscure key generalities, and to that end we consider $v_{F,0} = v_{F,\pi}=v_{F}$ a desirable simplification of this problem.  

Calculations by Killi, Wei, Affleck, and Paramekanti indicated that for interacting bilayer graphene systems, the interaction is dominated by two density-density interactions~\cite{affleck}:

\begin{eqnarray}
V_{+} &=& u_{+}(n_{0\uparrow} + n_{0\downarrow} + n_{\pi\uparrow} + n_{\pi\downarrow})^2 \nonumber \\  
V_{-} &=& u_{-}(n_{0\uparrow} + n_{0\downarrow} - n_{\pi\uparrow} - n_{\pi\downarrow})^2. 
\label{bosonInts}
\end{eqnarray}

$V_{+}$ is the usual two-body forward scattering term which leads to Luttinger liquid physics and $V_{-}$ is a new one which breaks the U(2) symmetry of electron distribution between the $0$ and $\pi$ bands.  Both can be effectively tuned by altering the strength of the perpendicular electric field, though for all reasonable numerical estimates, Ref.~\onlinecite{affleck} found $u_{-}<u_{+}$ and $u_{-}$ harder to tune, which is sensible as only $V_{+}$ contains contributions from the long-range part of the Coulomb interaction.  

Other density-density interactions, specifically those which affect electron spin, should be small in practice.  In the absence of an external magnetic field, and given the weak spin-orbit interaction in graphene, electron spin should remain SU(2)-symmetric.  In this limit, the electron spin sectors of this problem should remain noninteracting and terms which lead to phenomena such as spin-density waves will be marginally irrelevant~\cite{Giamarchi}. 

Returning to our Hamiltonian, we can consider a single domain wall by bosonizing,

\begin{equation}
\psi_{a\sigma,i} = \frac{1}{\sqrt{2\pi x_{c}}}e^{i\phi_{a\sigma,i}}
\end{equation}

where $a=0,\pi$; $\sigma=\uparrow,\downarrow$; $i=in,\ out$; and $x_{c}$ is the short wavelength cutoff. The bosonic fields $\phi_{a\sigma,i}$ obey the commutation algebra:

\begin{equation}
\left[\phi_{a\sigma,i}(x),\phi_{b\sigma',j}(y)\right]=i\pi\delta_{ab}\delta_{\sigma\sigma'}\tau^{z}_{ij}\sgn(x-y).
\label{OGcommutator}
\end{equation}

Under this transformation, the bare Hamiltonian and interactions become:

\begin{eqnarray}
\mathcal{H}_{0} &=& \frac{\hbar v_{F}}{4\pi}\sum_{\sigma=\uparrow,\downarrow}\big[(\partial_{x}\phi_{0\sigma,in})^{2} +(\partial_{x}\phi_{0\sigma,out})^{2} \nonumber \\
&+& (\partial_{x}\phi_{\pi\sigma,in})^{2} + (\partial_{x}\phi_{\pi\sigma,out})^{2} \big] \nonumber \\
V_{\pm} &=& \frac{\hbar v_{F}}{8\pi}\lambda_{\pm}\sum_{\sigma=\uparrow,\downarrow}\big[(\partial_{x}\phi_{0\sigma,in} - \partial_{x}\phi_{0\sigma,out}) \nonumber \\
&\pm& (\partial_{x}\phi_{\pi\sigma,in} - \partial_{x}\phi_{\pi\sigma,out})\big]^{2}  
\label{Hint}
\end{eqnarray}

where $\lambda_{\pm}=u_{\pm}/\pi\hbar v_{F}$.  The interacting Hamiltonian can be simplified by the sum/difference changes of basis:

\begin{eqnarray}
\phi_{\pm\sigma,i} &=& \phi_{0\sigma,i}\pm\phi_{\pi\sigma,i} \nonumber \\
\phi_{\pm c/s,i} &=& \phi_{\pm\uparrow,i} \pm \phi_{\pm\downarrow,i}
\label{Defs}
\end{eqnarray}

where the $c$ and $s$ sectors are charge and spin respectively.  In this basis, all of the interactions are in the charge sector and, as motivated earlier in this section, the spin sector is noninteracting:

\begin{eqnarray}
\mathcal{H}_{0} &=& \mathcal{H}_{+c} + \mathcal{H}_{-c} + \mathcal{H}_{+s} + \mathcal{H}_{-s} \nonumber \\
\mathcal{H}_{\pm c/s} &=& \frac{\hbar v_{F}}{8\pi}\left[(\partial_{x}\phi_{\pm c/s,in})^{2} + (\partial_{x}\phi_{\pm c/s,out})^{2}\right] \nonumber \\
V_{\pm} &=& \frac{\hbar v_{F}}{8\pi}\lambda_{\pm c}\left[\partial_{x}\phi_{\pm c,in} - \partial_{x}\phi_{\pm,out}\right]^{2}.
\end{eqnarray}

The plus and minus charge sectors of $\mathcal{H}_{0}$ are then each renormalized by only $V_{+/-}$ respectively, encouraging us to express the interaction parameter $g_{+/-}$ separately for each charge sector.  

Therefore we can write down the interacting Hamiltonian for each charge sector, $\mathcal{H}_{\pm c,int}=\mathcal{H}_{\pm c} + V_{\pm}$. Diagonalizing this Hamiltonian, the definition of the Luttinger parameters $g_{\pm}$ arises naturally.  The change of basis

\begin{equation}
\left(\begin{array}{c}
	\phi_{\pm c,in} \\
	\phi_{\pm c,out}
\end{array}\right) = \frac{1}{2g_{\pm}}\left(\begin{array}{cc}
	1 + g_{\pm} & 1 - g_{\pm} \\
	1 - g_{\pm} & 1 + g_{\pm}
\end{array}\right)\left(\begin{array}{c}
	\tilde{\phi}_{\pm c,in} \\
	\tilde{\phi}_{\pm c,out}
\end{array}\right)
\end{equation}

returns our interacting Hamiltonian to the form of one for non-interacting chiral bosons

\begin{eqnarray}
\mathcal{H}_{int} &=& \mathcal{H}_{+c,int} + \mathcal{H}_{-c,int} + \mathcal{H}_{+s} + \mathcal{H}_{-s}\nonumber \\
\mathcal{H}_{\pm c,int} &=& \frac{\hbar v_{\pm}}{8\pi g_{\pm}}\left[(\partial_{x}\tilde{\phi}_{\pm c,in})^{2} + (\partial_{x}\tilde{\phi}_{\pm c,out})^{2}\right]
\end{eqnarray}

where 

\begin{equation}
v_{\pm}=v_{F}\sqrt{1+2\lambda_{\pm}},\ g_{\pm} = \frac{1}{\sqrt{1+2\lambda_{\pm}}}.
\end{equation}

In the new basis, the charge fields $\tilde{\phi}_{\pm c,i}$ obey the commutation relation:

\begin{equation}
\left[\tilde{\phi}_{u c,i}(x),\tilde{\phi}_{v c,j}(y)\right]=i\pi g_{u}\delta_{uv}\tau^{z}_{ij}\sgn(x-y)
\end{equation}

where $u,v=+,-$; $i=in,\ out$; and we note that the noninteracting spin sector fields still obey this commutation relation with $g_{\pm}=1$.   

As in Refs.~\onlinecite{houkim,strom,teokane}, the tunneling density of states for a single edge $\rho(E)\propto E^{\alpha_{T}}$ is controlled by the interactions.  However here, unlike in the QSH case, the critical exponent is a function of two Luttinger parameters, such that in agreement with Ref.~\onlinecite{affleck}, 

\begin{equation}
\alpha_{T} = \frac{1}{8}(g_{+} + g_{-} + 1/g_{+} + 1/g_{-}) - \frac{1}{2}.
\end{equation}

From an experimental perspective, measuring this critical exponent for the tunneling conductance would be a valuable first step in confirming the Luttinger liquid physics of these bilayer graphene domain wall states.  

\subsection{Four-Terminal Geometry}
\label{sec:fourterminal}  

\begin{figure}
\centering
\includegraphics[width=3.5in]{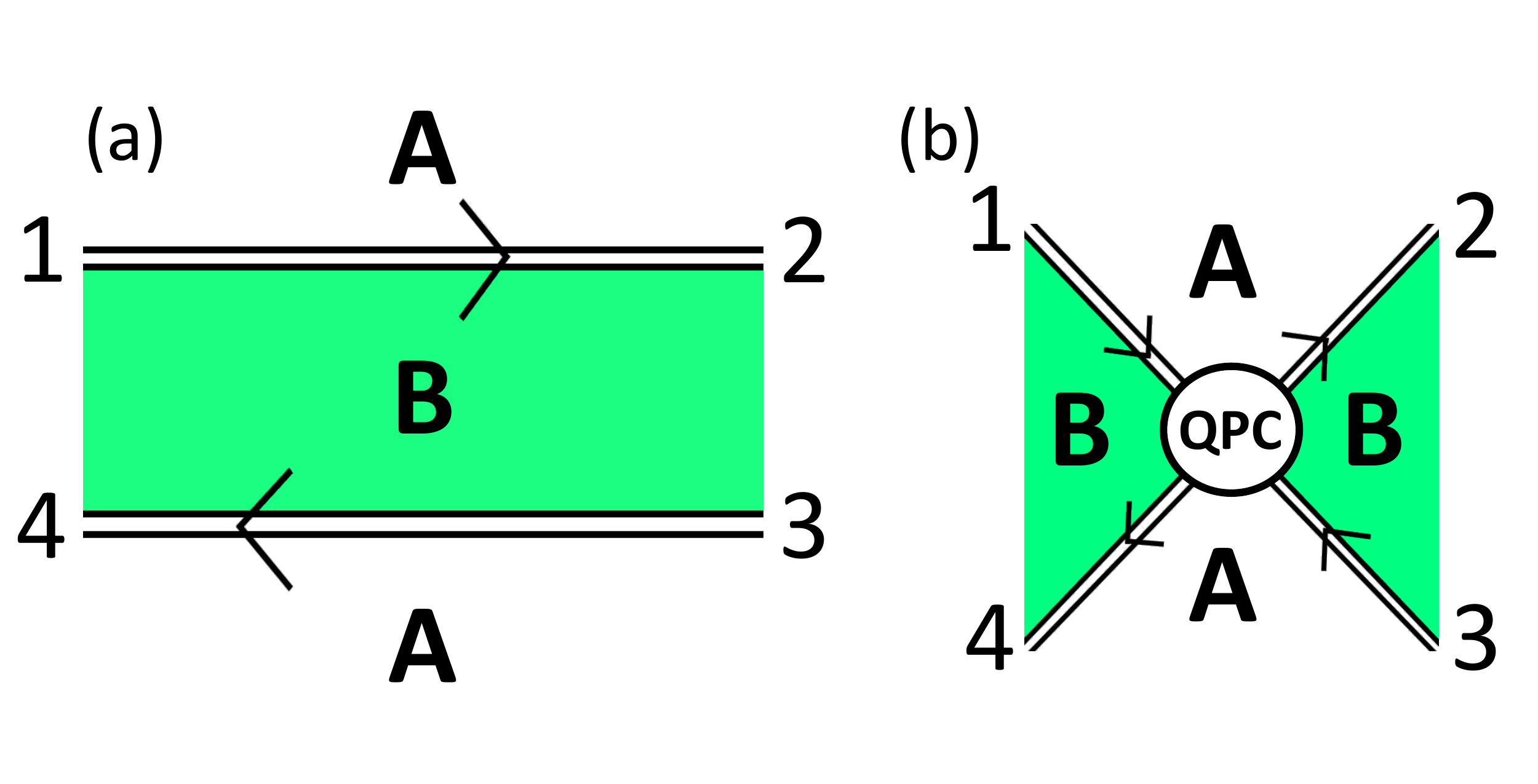}
\caption{(a) Two parallel domain walls in bilayer graphene can be created by varying either the interlayer stacking or the perpendicular field direction between regions {\bf A} and {\bf B}.  (b) Distorting region {\bf B} such that the walls approach each other results in the equivalent of a Quantum Point Contact (QPC) for the domain wall modes.  The numbers $1-4$ are lead indexes and the two modes displayed for each domain wall are those at $0$ and at $\pi$.  All of the modes shown here are at valley $K$; a counterpropagating set of modes exists at $K'$ and is related by time-reversal symmetry.  Including electron spin, there are 4 modes in each valley in each domain wall, for a total of 16 modes to consider for this QPC structure.}
\label{FigQPC}
\end{figure}

A pattern of two domain walls which pass nearly by each other can be formed by varying either electric field direction or interlayer stacking twice (Fig.~\ref{FigQPC}a).  If we distort the central region of this picture, we could imagine bringing the two domain walls so close that tunneling between them becomes significant.  In this case, the two domain walls have formed the equivalent of a quantum point contact (Fig.~\ref{FigQPC}b).  Mapping our work in~\ref{sec:domainwalls} onto a spinful Luttinger liquid for the two-wall system, we deduce in this section, for all interaction regimes, the interaction strengths at which many-body tunneling processes become relevant and alter the junction's charge and valley conductances.  

While each lead retains the properties and Luttinger parameters from~\ref{sec:model} individually, we will find it convenient to limit the usage of the $\tilde{\phi}_{\pm}$ basis to the treatment of isolated domain walls and adopt a new basis with effective charge and valley sectors.  The charge sector which arises here is a new degree of freedom and comes from a rotation of the indexes for $K$ and $K'$ and propagation direction.  We will label \emph{these} charge and valley sectors $\rho$ and $v$ respectively to differentiate them from the charge and spin sectors which arose in~\ref{sec:domainwalls}, which are labeled $c$ and $s$ respectively.

The two domain walls in Fig.~\ref{FigQPC} have opposite helicties, due to being on the top (bottom) of the central region.  We can define, for the interacting system, fields labeled by sum/difference, charge/spin, direction, and valley ($K,K'$):

\begin{eqnarray}
\phi_{\pm c/s,RK} &=& \phi_{\pm c/s,in,1}(-x)\Theta(-x) + \phi_{\pm c/s,out,2}(x)\Theta(x) \nonumber \\
\phi_{\pm c/s,LK'} &=& \phi_{\pm c/s,out,1}(-x)\Theta(-x) + \phi_{\pm c/s,in,2}(x)\Theta(x) \nonumber \\
\phi_{\pm c/s,LK} &=& \phi_{\pm c/s,in,3}(x)\Theta(x) + \phi_{\pm c/s,out,4}(-x)\Theta(-x) \nonumber \\
\phi_{\pm c/s,RK'} &=& \phi_{\pm c/s,out,3}(x)\Theta(x) + \phi_{\pm c/s,in,4}(-x)\Theta(-x)\nonumber \\ 
\end{eqnarray}

where $\Theta(x)$ is the Heaviside step function and the indexes $1-4$ on the noninteracting $\phi_{\pm}$ refer to the individual lead in Figure~\ref{FigQPC} with which they are associated.  The intersection of the two domain walls occurs at $x=0$ such that our theory consists of four, isolated domain walls everywhere except at that point.  Assuming that the interaction strength is controlled globally such that each lead has the same values of $\lambda_{\pm}$, we can create a $\rho/v$ basis for each $+/-$ sector of the combined Hamiltonian of the two domain walls $\mathcal{H}_{int}=\sum_{i=1}^{4}\sum_{\alpha=c,s}(\mathcal{H}_{+\alpha,int}^{i}+\mathcal{H}_{-\alpha,int}^{i})$:

\begin{eqnarray}
\phi_{\pm c/s,RK} &=& \frac{1}{2}\left[\phi_{\pm c/s,\rho} + \phi_{\pm c/s,v} + \theta_{\pm c/s,\rho} + \theta_{\pm c/s,v}\right] \nonumber \\
\phi_{\pm c/s,LK} &=& \frac{1}{2}\left[\phi_{\pm c/s,\rho} + \phi_{\pm c/s,v} - \theta_{\pm c/s,\rho} - \theta_{\pm c/s,v}\right] \nonumber \\
\phi_{\pm c/s,RK'} &=& \frac{1}{2}\left[\phi_{\pm c/s,\rho} - \phi_{\pm c/s,v} + \theta_{\pm c/s,\rho} - \theta_{\pm c/s,v}\right] \nonumber \\
\phi_{\pm c/s,LK'} &=& \frac{1}{2}\left[\phi_{\pm c/s,\rho} - \phi_{\pm c/s,v} - \theta_{\pm c/s,\rho} + \theta_{\pm c/s,v}\right] \nonumber \\ 
\end{eqnarray}

where again $c,s$ are the charge and spin sectors which resulted from rotating the indexes for $\uparrow,\downarrow$ and $\rho,v$ are the indexes which have, along with the choice of $\phi,\theta$, resulted from rotating indexes for propagation direction ($R,L$) and valley ($K,K'$).  The new fields are governed by the modified commutation relation:

\begin{equation}
\left[\theta_{u\alpha i}(x),\phi_{v\beta j}(y)\right]=2\pi i\delta_{uv}\delta_{\alpha\beta}\delta_{ij}\Theta(x-y)
\end{equation}

where $u,v=+,-$; $\alpha,\beta = c,s$; and $i,j=\rho,v$.   

This unitary rotation of the variables effectively changes the sign of the interaction ``cross-term'' individually for the choice of $\phi$, $\theta$, $\rho$, and $v$ within the $c$ sector:

\begin{eqnarray}
\mathcal{H}_{\pm c,int} &=& \frac{\hbar v_{F}}{8\pi}\bigg\{(1+\lambda_{\pm})\bigg[(\partial_{x}\phi_{\pm c\rho})^{2} + (\partial_{x}\phi_{\pm cv})^{2} \nonumber \\
&+& (\partial_{x}\theta_{\pm c\rho})^{2} + (\partial_{x}\theta_{\pm cv})^{2}\bigg] \nonumber \\
&-&\lambda_{\pm}\bigg[(\partial_{x}\phi_{\pm c\rho})(\partial_{x}\phi_{\pm c\rho}) - (\partial_{x}\phi_{\pm cv})(\partial_{x}\phi_{\pm cv}) \nonumber \\
&-& (\partial_{x}\theta_{\pm c\rho})(\partial_{x}\theta_{\pm c\rho}) + (\partial_{x}\theta_{\pm cv})(\partial_{x}\theta_{\pm cv})\bigg]\bigg\}.\nonumber \\
\label{spinful}
\end{eqnarray}

The previous equation, though diagonal, was left unsimplified and in the form of Eq.~\ref{Hint} such that by the same logic as in~\ref{sec:domainwalls}, the form of the simplified diagonalized Hamiltonian, as well as the interactions, can just be read off:

\begin{eqnarray}
\mathcal{H}_{\pm,int} &=& \frac{\hbar v_{\pm}}{8\pi}\sum_{\alpha=c,s}\sum_{a=\rho,v}g_{\pm\alpha a}(\partial_{x}\phi_{\pm\alpha a})^{2} + \frac{1}{g_{\pm\alpha a}}(\partial_{x}\theta_{\pm\alpha a})^{2} \nonumber \\
g_{\pm c\rho} &=& g_{\pm},\ \ g_{\pm cv} = 1/g_{\pm},\ \ g_{\pm s\rho}=g_{\pm sv}=1
\label{spinfulLL}
\end{eqnarray}

such that $\mathcal{H}_{int}$ now has the form of a spinful Luttinger liquid.  In this basis, both the interacting and noninteracting Hamiltonians are diagonal and so the transformation between the interacting and noninteracting $\theta/\phi$ requires just a simple rescaling by the interaction parameter for each sector.  

For this geometry, one can probe experimentally by measuring the current $I_{i}$ at one of the leads in response to an applied voltage on another lead $V_{j}$ such that a $4\times 4$ conductance matrix characterizes the system,

\begin{equation}
I_{i}=G_{ij}V_{j}
\end{equation}

where $i=1-4$ is a lead index.  In the presence of time-reversal and valley symmetries, the number of independent or nonzero parameters in $G_{ij}$ is greatly reduced, as described in detail in the appendix of Ref.~\onlinecite{teokane}.  For this system, we can then consider a reduced set of voltages and currents:

\begin{equation}
\left(\begin{array}{c}
I_{X} \\
I_{Y}
\end{array}\right)=\left(\begin{array}{cc}
G_{XX} & G_{XY} \\
G_{YX} & G_{YY}
\end{array}\right)\left(\begin{array}{c}
V_{X} \\
V_{Y}
\end{array}\right)
\end{equation}

where $I_{X}=I_{1}+I_{4}$ is the left-to-right current and  $I_{Y}=I_{1}+I_{2}$ is the top-to-bottom current.  $V_{X}$ and $V_{Y}$ are similarly defined such that $V_{X}$ is a bias of leads $1$ and $4$ relative to leads $2$ and $3$ and $V_{Y}$ is a bias of leads $1$ and $2$ relative to leads $3$ and $4$.  Therefore $G_{XX}$ and $G_{YY}$ are the two-terminal conductances measured left-to-right and top-to-bottom respectively.  $G_{XY}=G_{YX}$ are skew conductances, equal as a consequence of time-reversal symmetry.  In the noninteracting model, this skew conductance is zero as a consequence of artificial spatial symmetries, such as mirror symmetry.  Though it may become nonzero under increased interaction strengths, the skew conductance is still negligible along the relevant directions which characterize transitions in this system~\cite{teokane}.  We can define a final current across the junction $I_{Z}=I_{1}+I_{3}$, which one can probe by applying a voltage $V_{Z}$ which biases leads $1$ and $3$ relative to leads $2$ and $4$, with a conductance 

\begin{equation}
I_{Z}=G_{ZZ}V_{Z}.
\end{equation}

If valley is conserved, then electrons cannot enter at lead $1$ and exit at lead $3$, implying that a measurement of an exactly quantized

\begin{equation}
G_{ZZ}=\frac{8e^{2}}{h}
\end{equation}

would be an experimental confirmation that valley-nonconserving disorder is absent and the system is appropriately described by the physics in this paper.  The factor of $N=8$ in the Landauer prediction $G=Ne^{2}/h$ comes from factors of 2 for band index ($0$ and $\pi$), electron spin degeneracy, and the two incoming leads at $K$ ($1$ and $3$).  

We can also, in a similar manner, characterize the valley conductance of the system in terms of left-to-right and top-to-bottom parameters $G^{V}_{XX}$ and $G^{V}_{YY}$.  Before the introduction of any interactions or tunneling operators, our system consists of two, left-to-right domain wall states and we consider it ``fully-open.''  For this system, $G_{XX}=8e^{2}/h$ and $G^{V}_{XX}\neq 0$ such that it is a left-to-right \emph{charge conductor, valley conductor}, which we will denote as the CC phase.  A $90^{\circ}$ rotation and relabeling (with regards to Fig.~\ref{FigQPC}) or the pinch-off inversion of this phase, for which $G_{XX}=G^{V}_{XX}=0$ and $G_{YY}=8e^{2}/h,\ G^{V}_{YY}\neq 0$, is considered ``fully pinched-off'' and is a left-to-right \emph{charge insulator, valley insulator}, which we denote as the II phase.   

With this framework established, we can examine perturbatively tunneling processes between the two adjacent domain walls which may lead to differing charge and valley conductances.  Using our bosonization work, we can examine the rescaling of the coupling strength for each process, noting the interaction regime in which it dominates the physics of the quantum point contact.  

\begin{figure}
\centering
\scalebox{0.39}{\includegraphics*{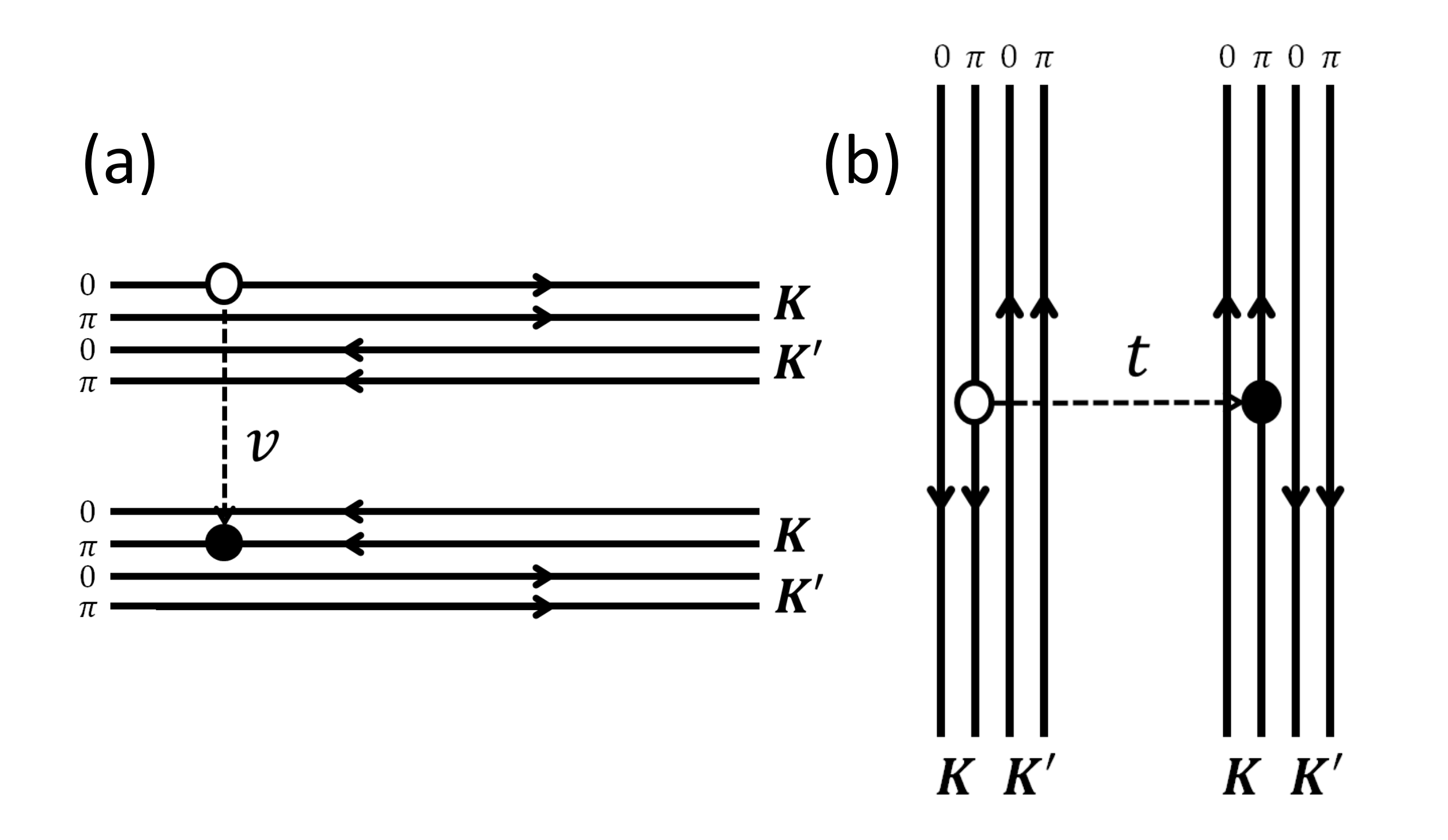}}
\caption{Schematic of the valley-preserving single-particle tunneling processes.  Many-body processes which conserve spin and valley can be constructed as products of these processes.  Among the processes which conserve valley, only spin-conserving processes can become relevant and destabilize the fully pinched-off II (t) and fully-open (v) CC phases, due to the nature of the scaling dimension calculation.  For each process about the charge and valley conducting phase (a), there is a dual process about the charge and valley insulating phase (b).  The diagram here depicts modes for only a single spin direction; the full QPC hosts an additional set of modes related by a spin flip.}  
\label{FigTV}
\end{figure}

Figure~\ref{FigTV} illustrates the single-particle valley-conserving processes which can exist in this system within a single spin channel.  In general, many-body tunneling processes will also be present and may become relevant.  These many-body tunneling processes can be considered products of single-electron-tunneling processes which, in the most general case, may or may not conserve spin or valley indexes.  However, restricting ourselves to the set of processes which conserve valley, it becomes apparent that the linear combinations of bosonized operators which can lead to relevant operators can only be achieved through products of single-particle processes which conserve spin.  Therefore, in the analysis of many-body processes which may become relevant and drive to phases with different conductance behavior, we can simply consider products of spin- and valley-conserving single-electron tunneling:

\begin{equation}
\mathcal{O}^{\alpha\beta}_{\sigma u} = \psi^{\dagger}_{\alpha\sigma Ru}\psi_{\beta\sigma Lu},\ V_{n-body} = v_{n}\prod_{i=1}^{n}\mathcal{O}_{i}+\ H.C. 
\end{equation}

where $\alpha,\beta=0,\pi$; $\sigma=\uparrow,\downarrow$; $u=K,K'$; $v_{n}$ is the coupling strength of the process; and $\mathcal{O}_{i}$ is an arbitrary valley- and spin-conserving single-particle tunneling process.  For the sake of condensing notation, tunneling from $\beta,L\rightarrow R,\alpha$ will be expressed as $O^{\alpha\beta\dagger}_{\sigma u}$.  Weak tunneling about the CC phase (Fig.~\ref{FigTV}a) is related to weak backscattering in the left-to-right direction, and is also dual to weak tunneling about the II phase (Fig.~\ref{FigTV}b).  Taking advantage of this duality, we will restrict our discussion to the set of $v$ operators which may destabilize the CC phase, noting that the $t$ tunneling operators about the II phase are related by a duality.  This duality is explored in greater detail in Ref.~\onlinecite{teokane}.

\begin{figure}
\centering
\includegraphics[width=3.5in]{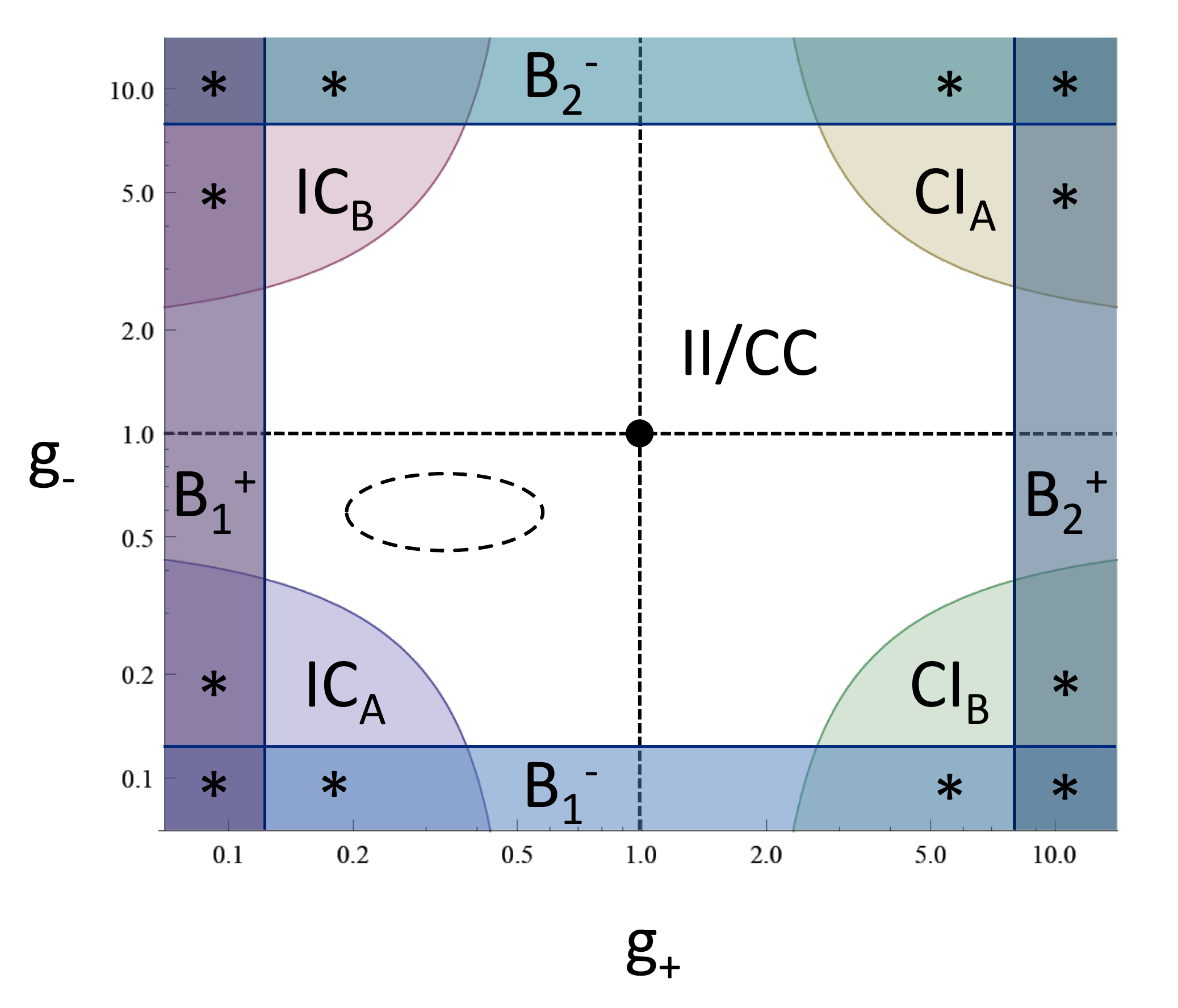}
\caption{The regions in interaction space for which tunneling processes become relevant ($\Delta< 1$ in Eqs.~(\ref{scalingD1}) and~(\ref{scalingD3})) and the fully open (CC) or pinched-off (II) junction phases are destabilized.  The central dot at $g_{+}=g_{-}=1$ is the noninteracting point and the dotted oval is the region of predicted accessible interaction strength in Ref.~\onlinecite{affleck} for a suspended sample.  The IC (CI) regions are characterized by four-body tunneling processes which transmit exclusively valley (charge) across the junction.  The B regions represent relevant eight-body tunneling processes which are charge insulating ($+$) or conducting ($-$) and differ by band-index and valley-transmission character.  Regions of overlap between boundaries, denoted with $*$, have multiple relevant operators at different orders and presumably more complicated behavior.  In the central region, the fully open (CC) or pinched-off (II) phases remain stable and the conductance is characterized by single-electron tunneling.}  
\label{FigPhase}
\end{figure}

The near-intersection of the two domain walls is a $0+1$-dimensional object, and therefore the coupling strength $v_{a}$ of a given tunneling process flows, to first order, as 

\begin{equation}
\frac{dv_{a}}{dl} = (1-\Delta(v_{a}))v_{a}
\end{equation}

where $\Delta(v_{a})$ is the scaling dimension of the tunneling process $V_{\alpha}$.  

To understand which operators may become relevant, we can first examine the single-electron tunneling processes (Fig.~\ref{FigTV}a):

\begin{equation}
V_{1} = v_{1}\sum_{\alpha\beta,\sigma,u}\mathcal{O}^{\alpha\beta}_{\sigma u} +\ H.C.
\end{equation}

where $\alpha,\beta=0,\pi$; $\sigma=\uparrow,\downarrow$; $u=K,K'$; and we have restricted ourselves to processes which preserve spin and valley.  

For single-particle tunneling,

\begin{equation}
\Delta(v_{1}) = \frac{1}{8}\left[g_{+} + \frac{1}{g_{+}} + g_{-} + \frac{1}{g_{-}} + 4\right]
\end{equation}

such that single-electron tunneling is always marginal or irrelevant ($\Delta(v_{1})\geq 1$) for all possible inter- and intra-band scattering processes.  In the nearly noninteracting regime, where the \emph{least irrelevant} operators are $V_{1}$, the strength of $v_{1}$ can be controlled by an external parameter, such as the gate voltage $V_{G}$ for a given set of interaction strengths $g_{\pm}$.  In this interaction regime, we know that the CC and II phases are stable and that at least one quantum critical point exists to mediate the transition between them.  In the subsequent section, Section \ref{sec:pinchoff}, we will use diagrammatic perturbation theory in the interactions about the CC phase to search for the set of possible intermediate phases and quantum critical points which characterize the single-electron-tunneling behavior of the junction.  

The factor of $4/8$ in $\Delta(v_{1})$ is due to $g_{\pm s\rho}=g_{\pm sv} = 1$ and will remain an obstacle to a process becoming relevant unless the operator in bosonized form only contains $c$-sector variables.  For higher-body tunneling, this factor is greater than or equal to $1$, and prevents any process which isn't pairwise spin conserving and invariant under arbitrary SU(2) spin rotations from becoming relevant.  One can view this as pairwise spin conservation allowing decomposition into products of $\mathcal{O}^{\alpha\beta}_{\sigma u}$ and SU(2) invariance providing the necessary linear combinations of $\sigma=\uparrow,\downarrow$ to isolate $c$ sector variables.  Restricting to processes which conserve valley and obey these spin index constraints,  the first operators to become relevant as interactions are increased are therefore a specific set of four- and eight-body tunneling processes.  

Figure~\ref{FigPhase} details the region in interaction space for which each class of many-body operators becomes relevant ($\Delta(v_{a})<1$).  When a process becomes relevant, the bosonized operators will become locked into the values which minimize the tunneling operator and become gapped out, altering the conductance of the junction.  As an applied voltage only couples to the total electron density, only relevant operators containing $\theta_{+c\rho}$ or $\phi_{+c\rho}$ can cause the junction to become charge insulating.  In the central region, all operators are marginal or irrelevant, though single-electron tunneling $V_{1}$ is only the least irrelevant operator close to the non-interacting point $g_{+}=g_{-}=1$.  

Four groups of four-body processes are present which can become relevant and drive to phases which are completely charge or valley insulating:

\begin{eqnarray}
V_{IC_{A}} &=& v_{IC_{A}}^{(1)}\left[\mathcal{O}^{00}_{\uparrow K}\mathcal{O}^{00}_{\uparrow K'}\mathcal{O}^{00}_{\downarrow K}\mathcal{O}^{00}_{\downarrow K'}+ 0\leftrightarrow\pi\right] \nonumber \\ 
&+& v_{IC_{A}}^{(2)}\left[\mathcal{O}^{0\pi}_{\uparrow K}\mathcal{O}^{\pi 0}_{\uparrow K'}\mathcal{O}^{0\pi}_{\downarrow K}\mathcal{O}^{\pi 0}_{\downarrow K'}+ 0\leftrightarrow\pi\right]+\ H.C. \nonumber \\ \nonumber \\
V_{IC_{B}} &=& v_{IC_{B}}^{(1)}\left[\mathcal{O}^{00}_{\uparrow K}\mathcal{O}^{\pi\pi}_{\uparrow K'}\mathcal{O}^{00}_{\downarrow K}\mathcal{O}^{\pi\pi}_{\downarrow K'}+ 0\leftrightarrow\pi\right] \nonumber \\
&+& v_{IC_{B}}^{(2)}\left[\mathcal{O}^{0\pi}_{\uparrow K}\mathcal{O}^{0\pi}_{\uparrow K'}\mathcal{O}^{0\pi}_{\downarrow K}\mathcal{O}^{0\pi}_{\downarrow K'}+ 0\leftrightarrow\pi\right] +\ H.C. \nonumber \\ \nonumber \\
V_{CI_{A}} &=& v_{CI_{A}}^{(1)}\left[\mathcal{O}^{00}_{\uparrow K}\mathcal{O}^{00\dagger}_{\uparrow K'}\mathcal{O}^{00}_{\downarrow K}\mathcal{O}^{00\dagger}_{\downarrow K'}+ 0\leftrightarrow\pi\right] \nonumber \\
&+& v_{CI_{A}}^{(2)}\left[\mathcal{O}^{0\pi}_{\uparrow K}\mathcal{O}^{\pi 0\dagger}_{\uparrow K'}\mathcal{O}^{0\pi}_{\downarrow K}\mathcal{O}^{\pi 0\dagger}_{\downarrow K'}+ 0\leftrightarrow\pi\right] +\ H.C. \nonumber \\ \nonumber \\
V_{CI_{B}} &=& v_{CI_{B}}^{(1)}\left[\mathcal{O}^{00}_{\uparrow K}\mathcal{O}^{\pi\pi\dagger}_{\uparrow K'}\mathcal{O}^{00}_{\downarrow K}\mathcal{O}^{\pi\pi\dagger}_{\downarrow K'}+ 0\leftrightarrow\pi\right] \nonumber \\
&+& v_{CI_{B}}^{(2)}\left[\mathcal{O}^{0\pi}_{\uparrow K}\mathcal{O}^{0\pi\dagger}_{\uparrow K'}\mathcal{O}^{0\pi}_{\downarrow K}\mathcal{O}^{0\pi\dagger}_{\downarrow K'}+ 0\leftrightarrow\pi\right] +\ H.C. \nonumber \\
\end{eqnarray}

or in bosonized form:

\begin{eqnarray}
V_{IC_{A}} &=& \cos(\theta_{+c\rho})\left[v^{(1)}_{IC_{A}}\cos(\theta_{-c\rho}) + v^{(2)}_{IC_{A}}\cos(\phi_{-cv})\right] \nonumber \\
V_{IC_{B}} &=& \cos(\theta_{+c\rho})\left[v^{(1)}_{IC_{B}}\cos(\theta_{-cv}) + v^{(2)}_{IC_{B}}\cos(\phi_{-c\rho})\right] \nonumber \\
V_{CI_{A}} &=& \cos(\theta_{+cv})\left[v^{(1)}_{CI_{A}}\cos(\theta_{-cv}) + v^{(2)}_{CI_{A}}\cos(\phi_{-c\rho})\right] \nonumber \\
V_{CI_{B}} &=& \cos(\theta_{+cv})\left[v^{(1)}_{CI_{B}}\cos(\theta_{-c\rho}) + v^{(2)}_{CI_{B}}\cos(\phi_{-cv})\right] \nonumber \\
\end{eqnarray}

where $v_{IC/CI}^{(1/2)}$ are the coupling constant strengths for the two different choices of interband scattering for each class of tunneling process and have absorbed factors of $2$ during bosonization and simplification.  The scaling dimensions for these couplings strengths are:

\begin{eqnarray}
\Delta(v^{(1)}_{IC_{A}}) &=& \Delta(v^{(2)}_{IC_{A}})=2g_{+} + 2g_{-} \nonumber \\ \nonumber \\
\Delta(v^{(1)}_{IC_{B}}) &=& \Delta(v^{(2)}_{IC_{B}})=2g_{+} + \frac{2}{g_{-}} \nonumber \\ \nonumber \\
\Delta(v^{(1)}_{CI_{A}}) &=& \Delta(v^{(2)}_{CI_{A}})=\frac{2}{g_{+}} + \frac{2}{g_{-}} \nonumber \\ \nonumber \\
\Delta(v^{(1)}_{CI_{B}}) &=& \Delta(v^{(2)}_{CI_{B}})=\frac{2}{g_{+}} + 2g_{-}.
\label{scalingD1} \nonumber \\ 
\end{eqnarray}

In the regions where the IC (CI) operators become relevant, the system will be charge (valley) insulating and valley (charge) conducting.  Expressed in terms of conductance elements, the IC (CI) phases will have $G_{XX}=G_{YY}=0$, $G^{V}_{XX}=G^{V}_{YY}\neq0$ ($G_{XX}=G_{YY}=8e^{2}/h$, $G^{V}_{XX}=G^{V}_{YY}=0$).  These phases are related to the charge and spin insulating phases for the topological insulator QPC~\cite{teokane}.  However, unlike those phases, the IC and CI phases in the bilayer graphene junction will still be transmitting in the spin sector as long as $g_{\pm s \rho/v}$ are relatively close to noninteracting.  

The remainder of the II/CC region is bounded by four eight-body processes.  Each one can be considered as a product of one of the terms in the previous four-body processes multiplied by a product of selected conjugates of itself as to isolate just a single charge-sector variable.  In bosonized variables, these eight-body tunneling operators are:

\begin{eqnarray}
V_{B_{1}^{\pm}} &=& v^{(1)}_{B_{1}^{\pm}}\cos(2\theta_{\pm c\rho}) + v^{(2)}_{B_{1}^{\pm}}\cos(2\phi_{\pm cv}) \nonumber \\
V_{B_{2}^{\pm}} &=& v^{(1)}_{B_{2}^{\pm}}\cos(2\theta_{\pm cv}) + v^{(2)}_{B_{2}^{\pm}}\cos(2\phi_{\pm c\rho})
\end{eqnarray}

The strengths of these operators have scaling dimensions:

\begin{eqnarray}
\Delta(v^{(1)}_{B_{1}^{\pm}}) = \Delta(v^{(2)}_{B_{1}^{\pm}}) = 8g_{\pm} \nonumber \\
\Delta(v^{(1)}_{B_{2}^{\pm}}) = \Delta(v^{(2)}_{B_{2}^{\pm}}) = \frac{8}{g_{\pm}}.
\label{scalingD3}
\end{eqnarray}

They are only relevant under extremely strong interactions ($g_{\pm}<1/8$ or $g_{\pm}>8$) and are charge insulating ($+$) or conducting ($-$).  

Higher-order many-body tunneling processes are of course also possibly relevant, however due to the nature of the scaling dimension calculation, they will become relevant at much larger values of interaction strength than the boundaries of the shaded regions in Figure~\ref{FigPhase}. 

\section{The Pinch-Off Transition}
\label{sec:pinchoff}

As demonstrated in the previous section, under weak interactions the junction is stable in either the open (CC) phase or the closed-off (II) phase, both of which are characterized by single-electron tunneling and are related to each other by both $90^{\circ}$ rotations and the pinch-off duality.  In this section, we expand perturbatively in the interactions about the CC fixed point in search of the quantum critical point(s) which characterize the CC$\leftrightarrow$II quantum phase transition.  In the process, we discover that, in addition to the $T_{0/\pi}=1/2$ critical point, which is expected as a consequence of the pinch-off duality, an additional family of intermediate critical points and phases are also present.  For each of the possible paths between the II and CC phases we derive the conductance signatures which characterize the low-temperature transitions as functions of the two interaction strengths and the external gate voltage.  First, we show how the general S-matrix characterizing the junction is renormalized by weak interactions, deriving a phase diagram and Renormalization Group (RG) in the case where scattering between the $0$ and $\pi$ bands is disallowed.  We then allow for interband scattering, as might be present in the case of relatively smooth disorder, and introduce an S-matrix parameterization incorporating the additional system parameters.  Using the results of an extensive renormalization group calculation, detailed in Appendix~\ref{appendix:stability}, we assert that the most general S-matrix flows back to one with small-momentum conservation.  The RG flow on this surface therefore contains all of the characteristic non-trivial quantum critical points for the pinch-off transition of this problem.  Finally, we derive the critical exponents and universal scaling functions for the two classes of conductance transitions, up to leading order in the interactions.  

\subsection{Non-Interacting Electrons}
\label{sec:noninteracting}

In the absence of interactions, tunneling through the junction can be characterized by an S-matrix restricted only by time-reversal symmetry and valley-index conservation

\begin{equation}
|\psi_{i,out}^{\alpha\sigma}\rangle = S_{ij}^{\alpha\beta}\delta^{\sigma\sigma '}|\psi_{j,in}^{\beta\sigma'}\rangle
\end{equation}

where $i,j$ are lead indexes $1-4$; $\sigma, \sigma '$ are spin indexes $\uparrow,\downarrow$; and $\alpha,\beta$ are band indexes $0$ and $\pi$.  Given the negligible spin-orbit coupling in graphene, we can consider here that the time-reversal operator $\tau = K$ leaves the spin sector unaffected, such that $\tau^{2} = +1$.  Therefore, time-reversal symmetry restricts that $S^{\alpha\beta}_{ij} = S^{\beta\alpha}_{ji}$. Additionally, to keep the domain wall states gapless, one must disallow scattering from $K$ to $K'$, which restricts elements of the S-matrix $S_{K}=S_{K'}^{T}$.  In this section, we will work the most general allowed S-matrix down to one which is characterized by parameters which have physical meaning.  Beginning with the modes in a single valley $K$:

\begin{equation}
S_{K}=\left(  
\begin{array}{cc}
     \mathbb{t} & \mathbb{r} \\
     -\mathbb{r}^{\dagger} & \mathbb{t}^{\dagger} 
  \end{array}\right)
\end{equation}

where the rows and columns of $S_{K}$ indicate scattering of the incoming modes with valley index $K$ (leads $1$ and $3$) to the outgoing ones (leads $2$ and $4$).  The matrices $\mathbb{r}$ and $\mathbb{t}$ live in the $2\times 2$ space of band indexes.  Identical copies of $S_{K}$ exist for the up and down spins.  The elements of $S_{K}$ are otherwise unconstrained if we allow scattering between the $0$ and $\pi$ bands such that $S_{K}$ is an arbitrary $U(4)$ matrix.  We can choose to parameterize 

\begin{equation}
S_{K} = \left(  
\begin{array}{cc}
     U_{1}^{\dagger} & 0 \\
     0 & U_{3}^{\dagger} 
  \end{array}\right)\left(  
\begin{array}{cc}
     \mathbb{t} & \mathbb{r} \\
     -\mathbb{r}^{\dagger} & \mathbb{t}^{\dagger}
  \end{array}\right) \left(  
\begin{array}{cc}
     U_{2} & 0 \\
     0 & U_{4}
  \end{array}\right)
\label{Smatrix}
\end{equation}

\begin{equation}
\mathbb{t} = \left(
 \begin{array}{cc}
    \sqrt{T_{0}} & 0 \\
    0 & \sqrt{T_{\pi}}
  \end{array}\right),\ \mathbb{r} = \left(
 \begin{array}{cc}
    \sqrt{1-T_{0}^{2}} & 0 \\
    0 & \sqrt{1-T_{\pi}^{2}}
  \end{array}\right) 
\end{equation}

where the $U_{i}$ ($U_{j}^{\dagger}$) are for now unconstrained U(2) matrices which characterize operations on the outgoing (incoming) electronic wavefunction at lead $i$ ($j$) and valley index $K$.  In this parameterization, we can choose the tunneling probabilities for each band to be real such that $T_{0/\pi}=|t_{0/\pi}|^2=\sqrt{1-|r_{0/\pi}|^{2}}$.  At this point, $S_{K}$ remains characterized by 16 free parameters.  Choosing to parameterize the $U_{i}$ in terms of Euler angles:

\begin{equation}
U_{i} = e^{i\phi_{i}\sigma^{z}}e^{i\theta_{i}\sigma^{y}}e^{i\alpha_{i}\sigma^{z}}e^{i\xi_i}.
\label{Umatrix}
\end{equation}

\begin{figure}
\centering
\includegraphics[width=3.5in]{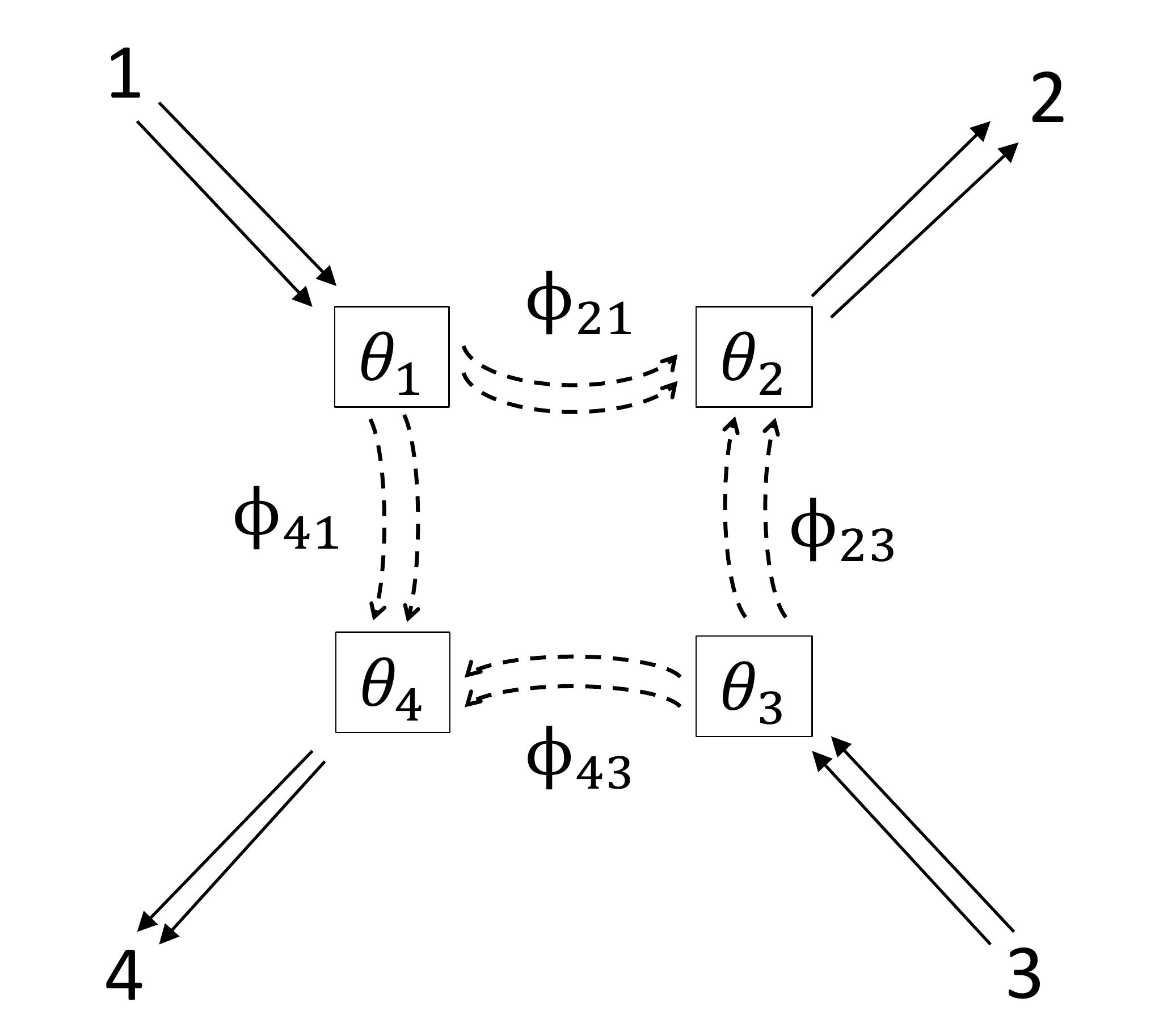}
\caption{The eight angular variables in the full formulation of the S-matrix, seven of which are linearly independent.  $\theta_{i}$ corresponds to the processes which break small-momentum conservation and scatter electrons between the $0$ and $\pi$ bands at lead $i$.  $\phi_{ij}=\phi_{i}-\phi_{j}$ corresponds to the scattering phase acquired tunneling from lead $j$ to lead $i$ at valley $K$.  While there are four independent $\theta_{i}$ angles, there are only three independent $\phi_{ij}$ as only the relative scattering phase matters, such that $\phi_{43} = \phi_{23} - \phi_{21} + \phi_{41}$.  We have only displayed modes at $K$; an additional copy of this picture exists for modes at $K'$, related by time-reversal symmetry.  This picture is valid for electrons with either spin, as our model system is SU(2) spin-invariant for all values of charge-sector interaction strengths.}  
\label{FigVars}
\end{figure}

Recognizing that $\alpha_i$ and $\xi_i$ are just U(1)$\times$U(1) transformations at each lead, we can gauge them out.  We are then left with nine variables which have physical significance.  The four $0\leftrightarrow\pi$ mixing angles $\theta_i$ correspond to the breaking of small-momentum-conservation at lead $i$, the three independent linear combinations of scattering phases $\phi_{ij}=\phi_{i}-\phi_{j}$ each correspond to the phase acquired for electrons scattering from lead $j$ to lead $i$ at valley index $K$, and the two real tunneling probabilities $T_{0}$ and $T_{\pi}$ characterize the extent to which the junction is pinched off for each band.  These seven linearly independent angular variables (illustrated in Fig.~\ref{FigVars}) and two tunneling probabilities completely span the space of the gauge-independent, time-reversal-symmetric, spin- and valley-index-conserving problem with S-matrix,

\begin{equation}
S=S_{K} \oplus S_{K'}=S_{K} \oplus S_{K}^{T}
\end{equation}

which has rows and columns characterized by lead indexes $i,j$.  This parameterization of the S-matrix in lead and band-index spaces is restated more explicitly in the beginning of Appendix~\ref{appendix:thetaphi}.  

In the subsequent sections, we will show how the surface where $\theta_i,\phi_{ij}=0$ is not just a welcome simplification of the problem, but also the surface which contains all of the characteristic quantum critical points and their single relevant eigenvectors.  

\subsection{S-Matrix Renormalization under Weak Interactions}
\label{sec:interactingRG}  

As discussed in~\ref{sec:domainwalls}, this system is characterized by two kinds of density-density interactions.  Here, we relate the S-matrix to the single-particle Green's function and then use diagrammatic perturbation theory in those two interactions to find the leading order corrections to the S-matrix and derive RG flow equations for our system parameters~\cite{matveev,teokane}.  We calculate the relevant physics on the high-symmetry surface where the interband scattering angles $\theta_i=0$.  In Appendix~\ref{appendix:stability}, we further calculate the RG flow for all nine S-matrix parameters, demonstrating that the angles $\theta_{i},\phi_{ij}$ either flow back to this surface or are marginal and trivial at the quantum critical points.  

\subsubsection{Constructing the S-Matrix Renormalization Group}
\label{sec:thetazero}

Scattering processes from one lead, band, and spin to another can be considered in terms of a single-electron thermal Green's function 

\begin{equation}
\mathbb{G}^{ab\alpha\beta,\sigma\sigma '}_{ij}(x,\tau,x',\tau ') = -i\left\langle T_{\tau}\left[\psi_{i}^{a\alpha\sigma}(x,\tau)\psi^{b\beta\sigma '\dagger}_{j}(x',\tau ')\right]\right\rangle
\end{equation}

where $T_{\tau}$ denotes imaginary time ordering and indexes $a,b=in,\ out$.  In the absence of interactions

\begin{equation}
\mathbb{G}^{\alpha\beta,\sigma\sigma '}_{ij}(z,z')=\frac{1}{2\pi i}\left(
\begin{array}{cc}
  \frac{\delta_{ij}\delta^{\alpha\beta}}{z-z'} & \frac{(S^{\beta\alpha}_{ji})^{*}}{z-\bar{z}'} \\
  \frac{S^{\alpha\beta}_{\ij}}{\bar{z}-z'} & \frac{\delta_{ij}\delta^{\alpha\beta}}{\bar{z}-\bar{z}'}
\end{array}\right)\delta^{\sigma\sigma '}
\end{equation}

where $z=\tau + ix$ and the rows and columns of $\mathbb{G}^{\alpha\beta}_{ij}$ are the $in/out$ indexes such that elements proportional to $S^{\alpha\beta}_{ij}$ correspond to $in\leftrightarrow out$.  Restricting ourselves to the interactions introduced in~\ref{sec:domainwalls} (Eq.~(\ref{bosonInts})), we can use diagrammatic perturbation theory to calculate the leading order corrections to $\mathbb{G}^{\alpha\beta,\sigma\sigma '}_{ij}$ in the presence of weak $u_{+/-}$, such that the Luttinger parameters $g_{+/-}=1 + \epsilon_{+/-}$ and terms are kept up to $\mathcal{O}(\epsilon_{+/-}^{2})$.

\begin{figure}
\centering
\includegraphics[width=3.5in]{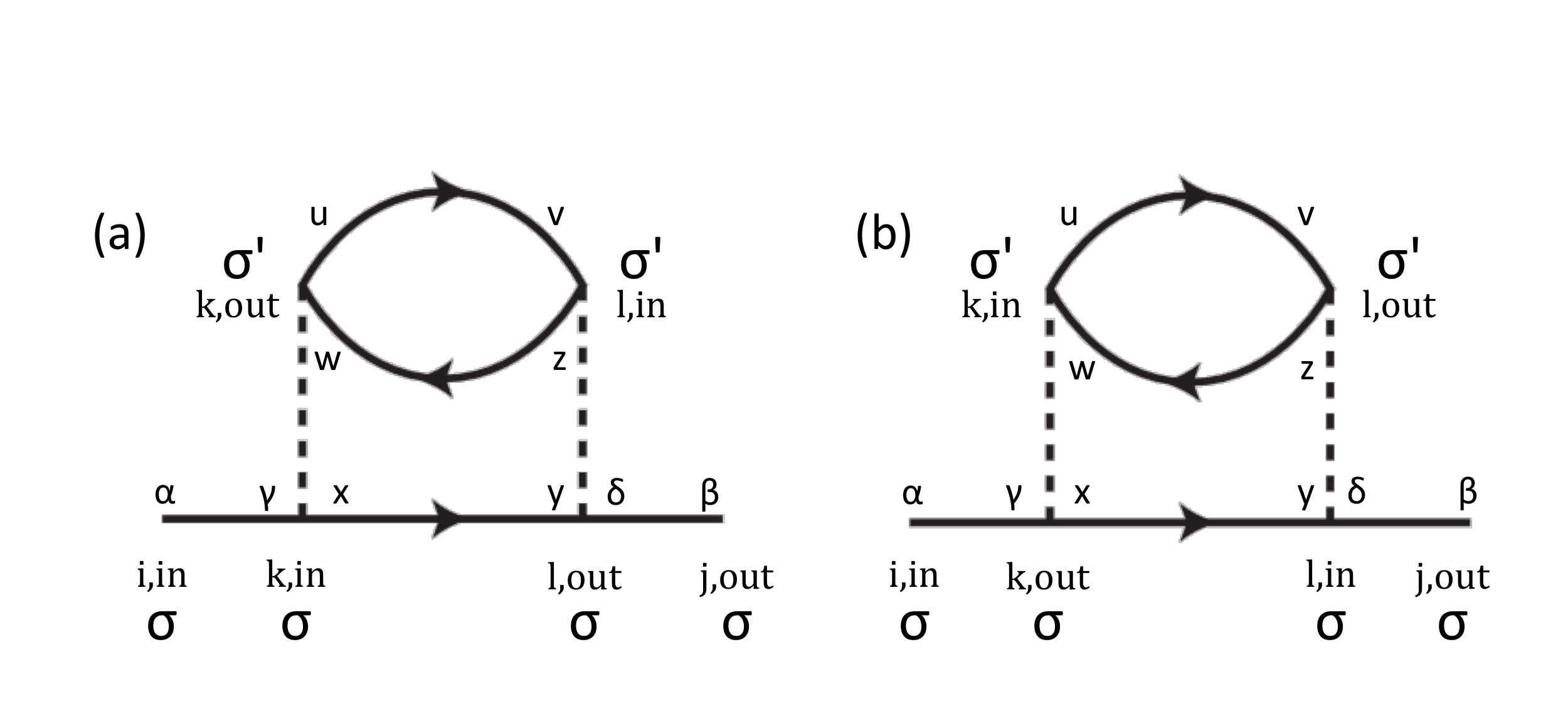}
\caption{The two non-zero, non-canceling diagrams for $\mathcal{O}(\epsilon_{+/-}^{2})$ perturbation theory. Note that $i-l$ are spatial lead indexes and $\alpha-\delta,u-z$ are band indexes which are summed at each vertex over $\epsilon^{\alpha\beta\gamma\delta}$.  $\sigma,\sigma '$ are spin indexes for which the $\delta^{\sigma\sigma '}$ within each Green's function has already been taken into account.  Even though the system is spin invariant, the spin index on the loop, here  $\sigma '$, must still be summed over its two values to calculate the physical RG flow.  Each diagram contributes a logarithmic correction to the S-matrix which, when renormalized, leads to a term in Eq.~(\ref{Sflow}).}
\label{FigDiagram}
\end{figure}

The one-loop corrections to the S-matrix are qualitatively similar to those in Ref.~\onlinecite{teokane}, with special care taken to properly sum at each vertex over the tensor structure of the interaction 
  
\begin{equation}
\epsilon^{\alpha\beta\gamma\delta}=\epsilon_{+}\delta^{\alpha\beta}\delta^{\gamma\delta}+\epsilon_{-}\sigma_{z}^{\alpha\beta}\sigma_{z}^{\gamma\delta}
\end{equation}

where $\alpha,\beta,\gamma,\delta$ are band indexes.  Illustrated in Fig.~\ref{FigDiagram}, only two diagrams contribute to the renormalization of the single-particle Green's function:

\begin{equation}
\mathbb{G}'^{out\ in,\alpha\beta,\sigma\sigma '} =\frac{1}{2\pi i}\frac{S'^{\alpha\beta}_{ij}}{\bar{z}-z'}\delta^{\sigma\sigma '}
\end{equation}

with 

\begin{widetext}
\begin{equation}
S'^{\alpha\beta}_{ij}=S^{\alpha\beta}_{ij} + 2\times\frac{1}{4}\log\frac{\Lambda}{E}\sum_{A,B=0,1}\epsilon_{A}\epsilon_{B}\left[\left(\sigma_{A}S_{ij}\sigma_{B}\right)^{\alpha\beta}\Tr[\sigma_{A}S^{\dagger}_{ji}\sigma_{B}S_{ji}] - \sum_{kl}\left(S_{ik}\sigma_{A}S^{\dagger}_{lk}\sigma_{B}S_{lj}\right)^{\alpha\beta}\Tr[\sigma_{A}S^{\dagger}_{lk}\sigma_{B}S_{lk}] \right]
\end{equation}
\end{widetext}

where $\Lambda$ and $E$ are the ultraviolet and infrared cutoffs respectively.  Traces refer to band index space, $\epsilon_{A/B}=\epsilon_{+/-}$, and $\sigma_{A/B}=\mathbb{1},\sigma_{z}$ ($\sigma_{0},\sigma_{1}$); tracing over the spin degree of freedom is implicit, which leads to the prefactor of $2$.  The two terms correspond to the diagrams (a) and (b) respectively in Fig.~\ref{FigDiagram}.  We can derive flow equations for the elements of $S^{\alpha\beta}_{ij}$ by rescaling the cutoff $\Lambda\rightarrow\Lambda e^{-l}$,

\begin{widetext}
\begin{equation}
\frac{dS^{\alpha\beta}_{ij}}{dl} = \frac{1}{2}\sum_{A,B=0,1}\epsilon_{A}\epsilon_{B}\left[\left(\sigma_{A}S_{ij}\sigma_{B}\right)^{\alpha\beta}\Tr[\sigma_{A}S^{\dagger}_{ji}\sigma_{B}S_{ji}] - \sum_{kl}\left(S_{ik}\sigma_{A}S^{\dagger}_{lk}\sigma_{B}S_{lj}\right)^{\alpha\beta}\Tr[\sigma_{A}S^{\dagger}_{lk}\sigma_{B}S_{lk}] \right].
\label{Sflow}
\end{equation}
\end{widetext}

One can immediately observe that this implies that $\theta_i=0$ is a fixed surface, since perturbative corrections to $S^{0\pi}_{ij}=0$ require multiplying by $S^{0\pi}_{ij}$ in the above equations.  We can thus, for now, simplify our focus to the fixed surface where $\theta_{i}=0$.  In Appendix~\ref{appendix:stability}, we calculate the stability of the quantum critical points on this surface for a general set of $\theta_{i}$ and $\phi_{ij}$, and demonstrate that for physical values of $g_{-}$ relative to $g_{+}$, the critical points are stable in all possible out-of-plane, $\tau$- and valley-symmetric directions. 

Using our parameterization of $S^{\alpha\beta}_{ij}$ on the $\theta_{i}=0$ surface, we can exploit the matrix structure of the diagrammatic perturbations to obtain flow equations for the transmission probabilities for each band

\begin{eqnarray}
\frac{dT_{0}}{dl} &=& -2T_{0}(1-T_{0})\bigg[(\epsilon_{+} + \epsilon_{-})^{2}(1-2T_{0})  \nonumber \\
&+& (\epsilon_{+} - \epsilon_{-})^{2}(1-2T_{\pi})\bigg] \nonumber \\
\frac{dT_{\pi}}{dl} &=& -2T_{\pi}(1-T_{\pi})\bigg[(\epsilon_{+} + \epsilon_{-})^{2}(1-2T_{\pi})  \nonumber \\ 
&+& (\epsilon_{+} - \epsilon_{-})^{2}(1-2T_{0})\bigg].
\label{flowEQ}
\end{eqnarray}

\begin{figure}
\centering
\includegraphics[width=3.5in]{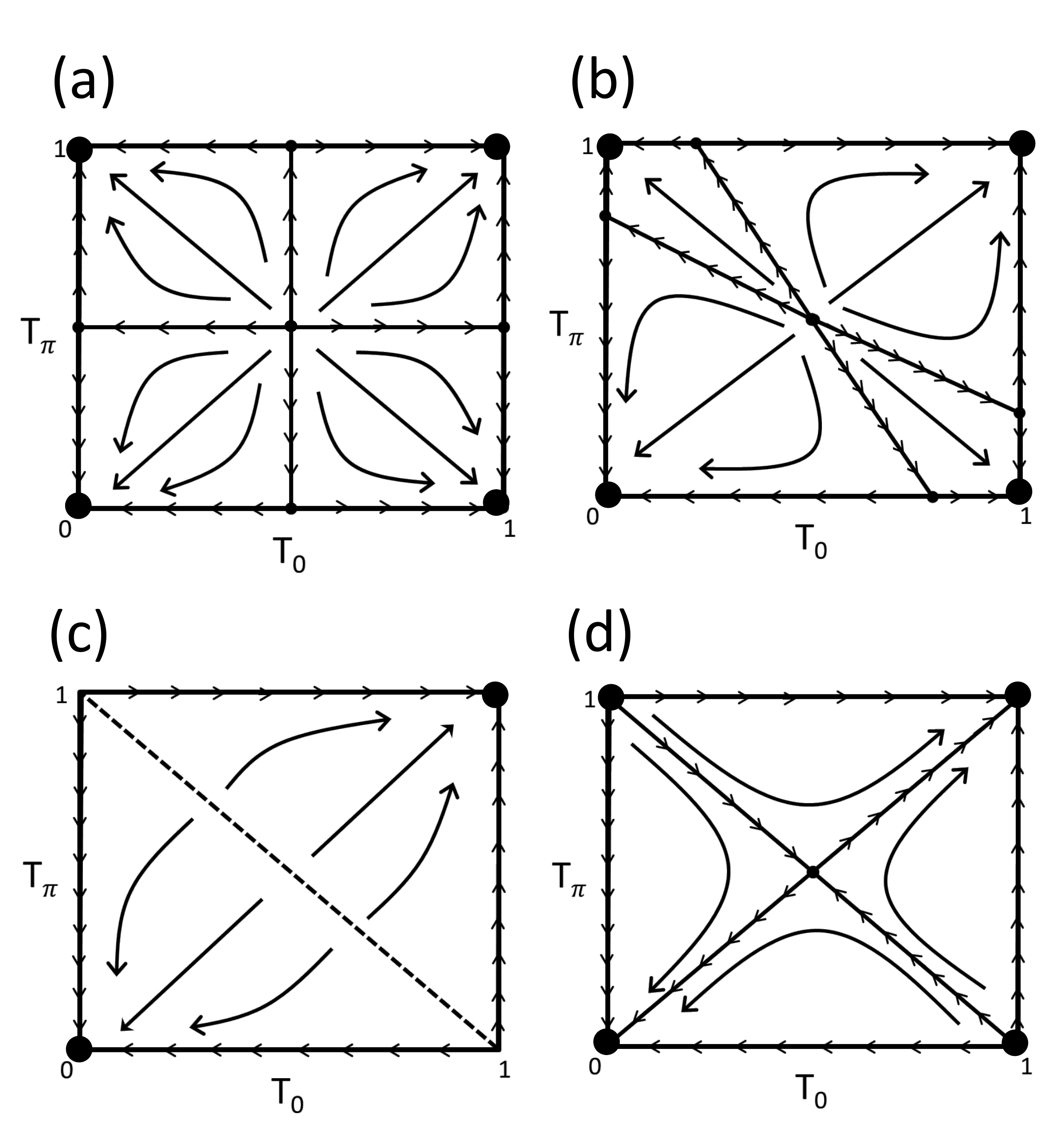}
\caption{RG flow of the variables $T_{0/\pi}$, which control the pinching off of the junction, calculated to quadratic order in the interactions $\epsilon_{+/-}$ (Eq.~(\ref{flowEQ})), panels (a)-(c).  Large circles correspond to stable fixed points and small circles indicate nontrivial quantum critical points.  The flow is controlled by the ratio of the interaction strengths $\epsilon_{-}/\epsilon_{+}$.  When $\epsilon_{-}/\epsilon_{+}=1$, the $0$ and $\pi$ bands are completely decoupled and each one behaves individually as a copy of the QSH problem in Ref.~\onlinecite{teokane} (a).   For $\epsilon_{-}/\epsilon_{+}>0$, a set of intermediate fixed points exists which allows the $0$ and $\pi$ bands to be pinched off independently (b).  When $\epsilon_{-}=0$, the quadratic theory predicts that a fixed line will exist $T_{0}+T_{\pi}=1$ (c).  Higher-order corrections about this line, calculated in Appendix~\ref{appendix:bosonization}, infer flow along it back to the central quantum critical point $T_{0}=T_{\pi}=1/2$ (d).}
\label{FigQuad}
\end{figure}

This system of equations obeys two key symmetries. First and foremost, like the QSH point contact described in Ref.~\onlinecite{teokane}, it is invariant under the pinch-off duality $T_{0/\pi}\leftrightarrow(1-T_{0/\pi})$, which we introduced in~\ref{sec:fourterminal}.  Eqs.~(\ref{flowEQ}) are also invariant under exchange of the band indexes $0$ and $\pi$.  In the context of our renormalization group calculation, $0$ and $\pi$ are arbitrary labels for the band degree of freedom, and so even though this system contains stable fixed points with broken band-index symmetry, the system of flow equations itself must be band-index symmetric.  Graphically, these two symmetries manifest themselves as two mirror symmetries in Fig.~\ref{FigQuad}: one about $T_{0}+T_{\pi}=1$ and the other about $T_{0}=T_{\pi}$.  

\subsubsection{Fixed Points and Renormalization Group Flow}
\label{section:qcps}

This system of flow equations can have as many as nine fixed points to quadratic order in the interactions (Fig.~\ref{FigQuad}).  The two corner fixed points at $T_{0}=T_{\pi}=0,1$ are stable for all values of $\epsilon_{-}$.  The central point at $T_{0}=T_{\pi}=1/2$ controls the transition between the CC and II corners and is related to the $T=1/2$ critical point in  the single-band TI QPC case~\cite{teokane}.  As in that case, the existence of this central quantum critical point is mandated by the pinch-off duality $T_{0/\pi}\leftrightarrow 1-T_{0/\pi}$.  The corner fixed points at $T_{0}=0,\ T_{\pi}=1$ and $T_{0}=1,\ T_{\pi}=0$ represent new, stable, single-electron-tunneling phases where \emph{only one of the bands} is pinched off.  We label these intermediate, mixed-band-character fixed points as ``M Phases''.   Transitions between the fully open or closed phases and these M phases are controlled by four fixed points which exist for $\epsilon_{-}/\epsilon_{+}>0$ (Figs.~\ref{FigQuad}a,~\ref{FigQuad}b).  For $\epsilon_{-}=\epsilon_{+}$, the $0$ and $\pi$ bands are completely decoupled and each one acts as a (spin-doubled) independent copy of the QSH problem in Ref.~\onlinecite{teokane} (Fig.~\ref{FigQuad}a).  When $\epsilon_{-}/\epsilon_{+}=0$, Eq.~(\ref{flowEQ}) predicts that all of the intermediate transitions and stable fixed points will collapse onto a fixed line at $T_{0}+T_{\pi}=1$ (Fig.~\ref{FigQuad}c).  This $\epsilon_{-}=0$ case represents a restoration of U(2) band-index symmetry locally at each lead under which the band indexes become trivial.  

We can examine this fixed line in greater detail by expanding upon our bosonization calculations from~\ref{sec:fourterminal}.  In Appendix~\ref{appendix:bosonization}, we increase the strength of $\pi\leftrightarrow\pi$ single-electron tunneling to drive from the CC phase towards an action about the M-phase corner fixed points for $g_{-}=1$.  Calculating higher-order correlation functions about this theory and expanding perturbatively again in the interactions, we discover additional fixed points:

\begin{equation}
1-T_{0}=T_{\pi}=\frac{48}{\pi^{2}}\frac{\epsilon_{-}}{\epsilon_{+}^{3}},\ T_{0}=1-T_{\pi}=\frac{48}{\pi^{2}}\frac{\epsilon_{-}}{\epsilon_{+}^{3}}
\end{equation}

where the second point is implied by the combination of mirror reflections about the pinch-off and band-index-exchange lines. Taking $\epsilon_{-}\rightarrow 0$, this theory exhibits flow back towards the central quantum critical point $T_{0}=T_{\pi}=1/2$ (Fig.~\ref{FigQuad}d).  The simplest assumption would be to postulate that, to lowest order, this flow continues away from the vicinity of the M points without additional fixed points appearing.  This implies that the fixed line at $\epsilon_{-}=0$ is simply an artifact of the $\mathcal{O}(\epsilon_{+/-}^{2})$ perturbation theory and that for extremely small $\epsilon_{-}/\epsilon_{+}$, the M phase is unstable and flow lines in that region point towards the central quantum critical point $T_{0}=T_{\pi}=1/2$.  From this information, we obtain Figure~\ref{FigRG}, a schematic which incorporates the quadratic-order RG flow and the higher-order corrections near the $T_{0}+T_{\pi}=1$ line.  As highlighted by the red and purple arrows in that figure, each of the two classes of nontrivial quantum critical points is characterized by only a single relevant direction in the $T_{0}-T_{\pi}$ plane.  All other out-of-plane $\theta_{i}$ and $\phi_{ij}$ directions are irrelevant or trivially marginal (Appendix~\ref{appendix:stability}).   

\subsubsection{Conductance Signatures and Universal Scaling Functions}
\label{sec:scaling}    

For each class of quantum critical point, we can, knowing that its only relevant eigenvector lies on the $\theta_{i}=0$ plane, use the quadratic-order flow equations to calculate the universal conductance scaling and critical exponents to leading order in the interactions.  

Returning to a discussion of reduced conductance matrixes from~\ref{sec:fourterminal}, we can express the left-to-right, two-terminal conductance $G_{XX}$ in terms of the S-matrix.  The elements of the four-terminal conductance $G$ in the lead basis are related to the S-matrix by:

\begin{equation}
G_{ij} = \frac{2e^{2}}{h}\Tr[\mathbb{1}-S^{\dagger}_{ij}S_{ij}]
\end{equation}

where $i,j$ are lead indexes $1-4$ such that the matrix

\begin{widetext}
\begin{equation}
G = \frac{2e^{2}}{h}\left(\begin{array}{cccc}
2 & -T_{+} & 0 & -(2-T_{+}) \\
-T_{+} & 2 & -(2-T_{+}) & 0 \\
0 & -(2-T_{+}) & 2 & -T_{+} \\
-(2-T_{+}) & 0 & -T_{+} & 2
\end{array}\right)
\end{equation}
\end{widetext}

\begin{figure}
\centering
\includegraphics[width=3.5in]{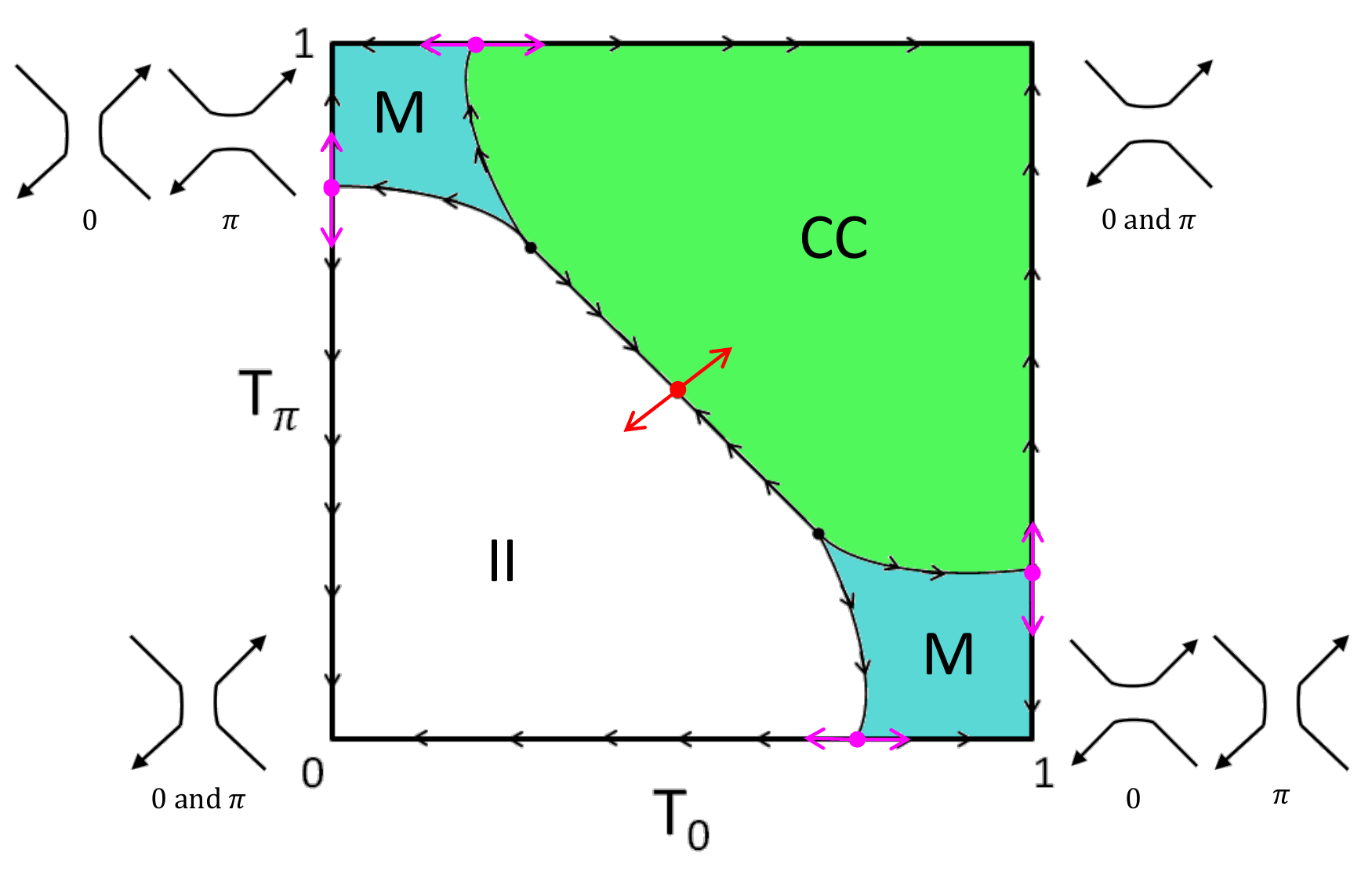}
\caption{A schematic phase diagram, in terms of left-to-right conductance, within the $T_{0}-T_{\pi}$ plane, combining information from Eq.~(\ref{flowEQ}) and Appendix~\ref{appendix:bosonization}.  There are two classes of quantum critical points.  The central point controls transitions between the fully open (CC) phase and the fully pinched-off (II) phase.  Four additional critical points on the edges control transitions between the CC/II phases and an intermediate mixed (M) phase in which the two bands have differing conductance contributions. The width of the M phase is $\mathcal{O}(\epsilon_{-}/\epsilon_{+})$.}
\label{FigRG}
\end{figure}

where $T_{\pm} = T_{0} \pm T_{\pi}$ and the factor of $2$ on the conductance is due to electron spin degeneracy.  The linear combinations of lead currents $I_{1-4}$ which give $I_{X,Y,Z}$ are a result of, in combination with the requirement $\sum_{i}I_{i} = 0$, 

\begin{equation}
\left(\begin{array}{c}
I_{X} \\
I_{Y} \\
I_{Z} 
\end{array}\right) = M^{T}\left(\begin{array}{c}
I_{1} \\
I_{2} \\
I_{3} \\
I_{4}
\end{array}\right)
\end{equation}

where

\begin{equation}
M = \frac{1}{2}\left(\begin{array}{ccc}
1 & 1 & 1 \\
-1 & 1 & -1 \\
-1 & -1 & 1 \\
1 & -1 & -1
\end{array}\right)
\end{equation}

such that

\begin{equation}
G^{XYZ} = M^{T}GM = \frac{4e^{2}}{h}\left(\begin{array}{ccc}
T_{+} & 0 & 0 \\
0 & 2-T_{+} & 0 \\
0 & 0 & 2
\end{array}\right).
\end{equation}

This confirms the result from~\ref{sec:fourterminal} that, for the valley-conserving problem, $G_{ZZ}$ is quantized to be $8e^{2}/h$ regardless of the junction state.  This reduction also confirms that, in terms of the S-matrix elements,

\begin{equation}
G_{XX} = \frac{4e^{2}}{h}(T_{0} + T_{\pi}) = \frac{8e^{2}}{h} - G_{YY}.
\label{2term}
\end{equation}  

With the structure of the conductance matrix established we can analyze the finite-temperature conductance transitions near each transition voltage $V_{G}^{*}$.  

\begin{figure}
\centering
\includegraphics[width=3.5in]{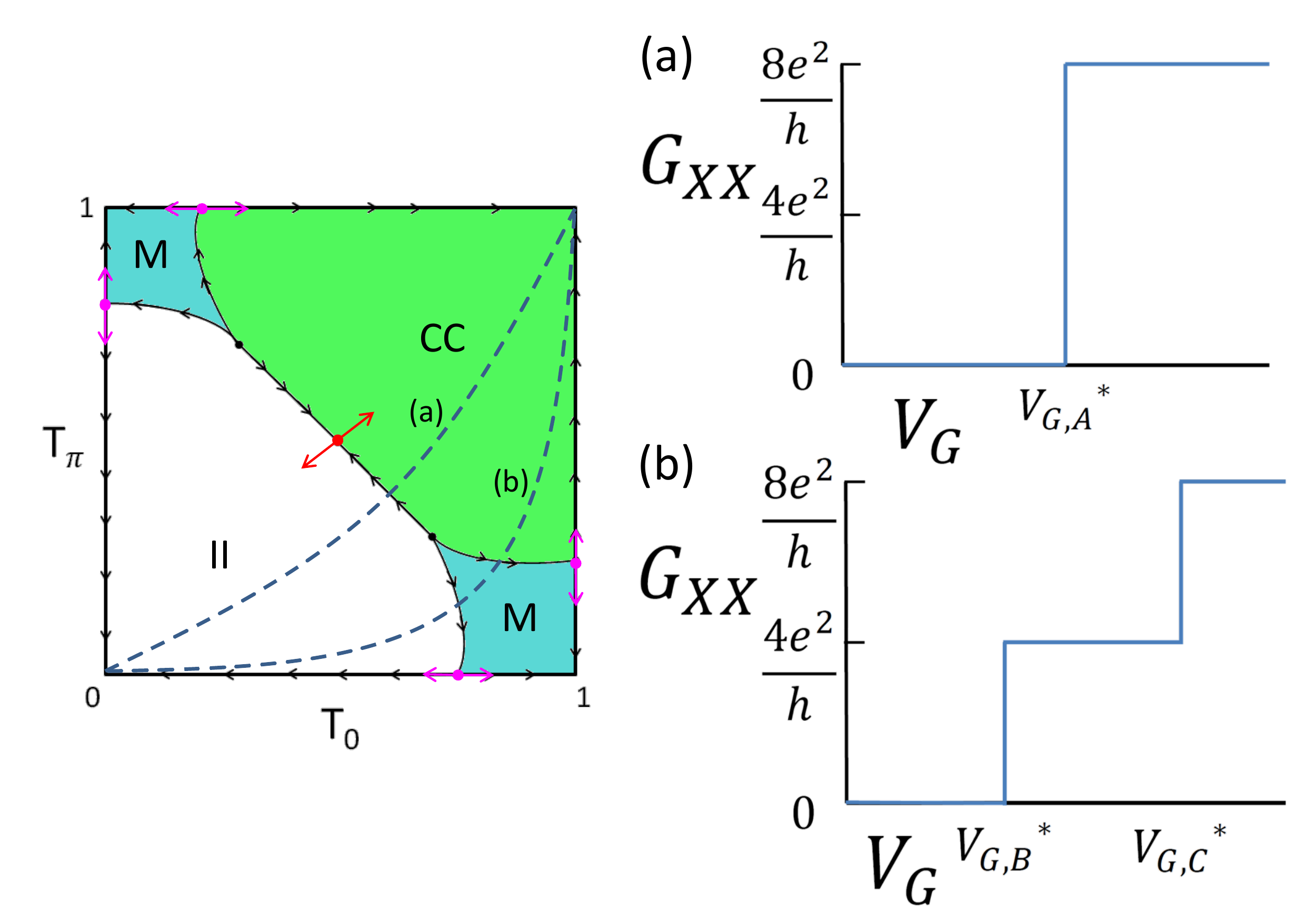}
\caption{A reproduction of Figure~\ref{FigRG} with dashed lines overlaid to indicate possible voltage curves.  As the gate voltage $V_{G}$ winds along a voltage curve, whose exact curvature is dictated by experimental specifics, it passes directly from the II to the CC region along a curve like (a) or indirectly, passing along the way through an intermediate M phase along a curve like (b).  At zero temperature, the left-to-right conductance $G_{XX}$ will therefore undergo a direct transition from $0\rightarrow 8e^{2}/h$ along (a) or one with an intermediate step up to $4e^{2}/h$ along (b).  This behavior motivates us to search for the finite-temperature scaling of these conductance transitions for $V_{G}\sim V_{G,A}^{*}$ (a), or for $V_{G}\sim V_{G,B}^{*}$ and $V_{G}\sim V_{G,C}^{*}$ (b).}
\label{FigZeroT}
\end{figure}

First, we will consider the direct II-CC phase transition, for which $G_{XX}$ scales from $0$ to $8e^{2}/h$.  We can write the conductance in its scaling form

\begin{equation}
G_{XX,A}(\Delta V_{G},T)=8\frac{e^{2}}{h}\mathcal{G}_{A}\left(c\frac{\Delta V_{G}}{T^{\alpha_{A}}}\right) 
\end{equation}

where $\Delta V_{G}=V_{G}-V_{G,A}^{*}$ in Fig.~\ref{FigZeroT}, $c$ is a non-universal constant, and the subscript $A$ on $\mathcal{G}$ and $\alpha$ denotes the direct quantum phase transition between the II and CC phases.  Observing that infinitesimal movement of $T_{-}$ away from $0$ is irrelevant (Fig.~\ref{FigRG}), we can set $T_{-}=0$ and characterize the conductance transition from II-CC with a single parameter,

\begin{equation}
\frac{dT_{+}}{dl} = -2(\epsilon_{+}^{2}+\epsilon_{-}^{2})T_{+}(2-T_{+})(1-T_{+}).
\end{equation}

This equation can be integrated to determine the crossover scaling function.  Taking $T_{+}=T^{0}_{+}$ at $l=0$,

\begin{equation}
\frac{T^{0}_{+}(2-T^{0}_{+})}{(1-T^{0}_{+})^{2}}e^{-4(\epsilon_{+}^{2}+\epsilon_{-}^{2})l}=\frac{T_{+}(2-T_{+})}{(1-T_{+})^{2}}
\label{tplusofl}
\end{equation}

where we have purposefully left $T_{+}(l)$ in its implicit form to provide a framework for the more mathematically complicated II-M transition analysis later in this section.  As one adjusts the gate voltage, $T^{0}_{+}$ passes through $1$ at $V_{G}=V_{G,A}^{*}$, such that near the transition $\Delta V_{G}\propto T^{0}_{+}-1$.  To determine the critical behavior, we can therefore expand $T^{0}_{+}$ around this value.  Finite temperature $T$ can be taken into consideration by cutting off the renormalization group flow $l$ at $\Lambda e^{-l} \propto T$.  Taking $\Delta V_{G},T\rightarrow 0$ at an arbitrary ratio, the previous equation can be rewritten 

\begin{equation}
\frac{1}{(2X)^{2}}=\frac{\mathcal{G}_{A}(1-\mathcal{G}_{A})}{(1-2\mathcal{G}_{A})^{2}}
\label{GAscaling}
\end{equation}

or explicitly inverted

\begin{equation}
\mathcal{G}_{A}(X) = \frac{1}{2}\left[1 + \frac{X}{\sqrt{1+X^{2}}}\right]
\end{equation}

where $T_{+}=2\mathcal{G}_{A}$ such that $X\propto \Delta V_{G}/T^{2(\epsilon_{+}^{2} + \epsilon_{-}^{2})}$.

\begin{figure}
\centering
\includegraphics[width=3.4in]{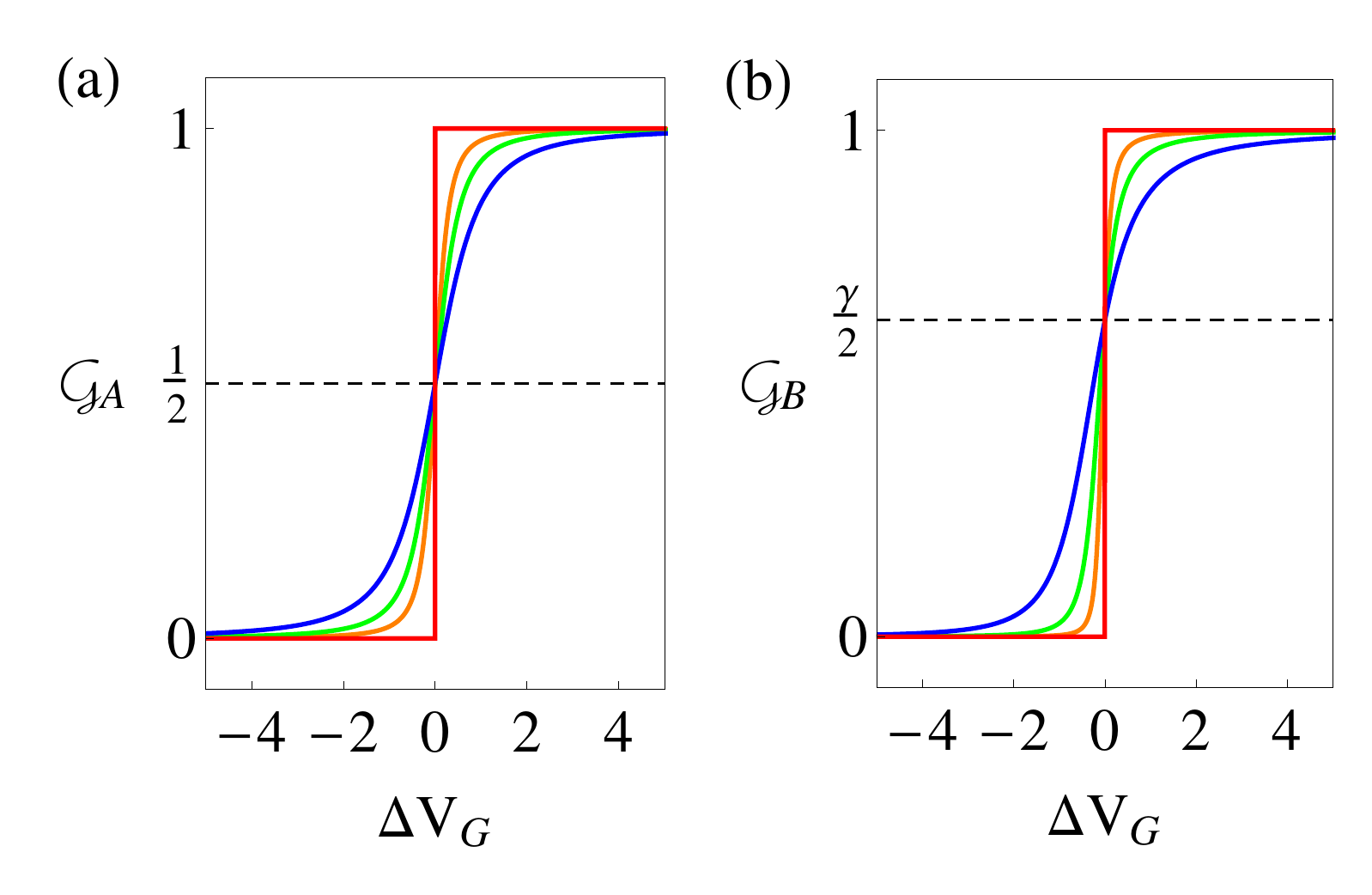}
\caption{The two classes of universal scaling functions as functions of external gate voltage: $\mathcal{G}_{A}$ describes the direct II-CC quantum phase transition and $\mathcal{G}_{B}$ describes the II-M transition, plotted in panels (a) and (b) respectively.  Here, we have plotted using $\epsilon_{+}=0.212$, $\epsilon_{-}=0.071$, such that $\gamma=1.25$.  The curves are plotted for increasing temperature, with the red, orange, green, and blue curves representing $T/c^{1/\alpha} = 0$, $10^{-5}$, $10^{-2.5}$, and $1\ V^{1/\alpha}$ respectively in equations~(\ref{GAscaling}) and~(\ref{GBscaling}).  Note that the crossover value of $\mathcal{G}_{A}$ is fixed to be $1/2$, whereas the crossover value for $\mathcal{G}_{B}$ is instead at $\gamma/2$, where $\gamma$ varies from $1$ to $2$ continuously as a function of interaction strength.}
\label{FigTemp}
\end{figure}

\begin{equation}
\alpha_{A}=2(\epsilon_{+}^{2} + \epsilon_{-}^{2})
\end{equation}

is the universal critical exponent for the II-CC quantum phase transition.  Figure~\ref{FigTemp} shows $\mathcal{G}_{A}$ at finite temperatures, noting that it collapses onto a step function at $T/c^{1/\alpha}=0$ and that it has a crossover point pinned at $\mathcal{G}_{A}=1/2$ for all values of interaction, making it identical to the T-R scaling function for weak interactions in the related Quantum Spin Hall problem~\cite{teokane}.  When $\epsilon_{-}=0$ and the only transitions are directly from II-CC, $\alpha_{A}=4\alpha_{QSH}$, where $\alpha_{QSH}$ is the T-R critical exponent in Ref.~\onlinecite{teokane}.  This factor of $4$ accounts for the fact that even though here only a single linear combination of indexes is being pinched off, the diagrams which contribute to the flow equations still live in a matrix space which is four times as large as that of the comparable Quantum Spin Hall problem.  

Obtaining the critical exponent and scaling function for the II-M quantum phase transition is procedurally identical.  As an example, we will choose the bottom right transition point in Fig.~\ref{FigZeroT} for our calculation, though that point is restricted to be the same as the one characterizing the finite $T_{\pi}$ II-M transition by band-index-exchange symmetry.  The quantum critical point is located at $T_{\pi}=0$, $T_{0} = \gamma/2$ where 

\begin{equation}
\gamma= 1 + \frac{(\epsilon_{-}-\epsilon_{+})^{2}}{(\epsilon_{-}+\epsilon_{+})^{2}}.
\end{equation} 

First and foremost, we can note that when $\gamma=2$, $\epsilon_{-}=0$ and there is no more available phase space (at quadratic order) for the II-M transition to exist.  Near this case, the M phase will exist in a vanishing area of phase space and most transitions will be controlled by the central quantum critical point in Figure~\ref{FigRG}.  However, in Appendix~\ref{appendix:stability} we have only analytically calculated the stability of the II-M quantum critical point to linear order in $\epsilon_{-}$, whereas our analysis of the critical behavior of the II-M transition is up to $\mathcal{O}(\epsilon_{-}^{2})$.  We believe it reasonable to assume that this stability extends up to quadratic order in the interactions such that these quantum critical points still describe the relevant physical transitions in this problem.  

For the II-M transition, the conductance jumps from $G_{XX}=0\rightarrow 4e^{2}/h$, with here the $T_{0}$ axis being the only relevant direction.  For this transition then, $G_{XX} = \frac{4e^{2}}{h}T_{0}$.  Expressing the conductance in its scaling form

\begin{equation}
G_{XX,B}(\Delta V_{G},T)=4\frac{e^{2}}{h}\mathcal{G}_{B}\left(c\frac{\Delta V_{G}}{T^{\alpha_{B}}}\right)
\end{equation}

where again $\Delta V_{G}$ is the external gate voltage, $c$ is a non-universal constant that may certainly differ from the $c$ in the II-CC transition, and the subscript $B$ denotes that transition between the II-M phases.  Taking $T_{\pi}=0$, the conductance transition from II-M is characterized by the flow of a single parameter,

\begin{equation}
\frac{dT_{0}}{dl} = -2\gamma(\epsilon_{+} + \epsilon_{-})^{2}T_{0}(1-T_{0})\left(1-\frac{2T_{0}}{\gamma}\right).
\end{equation}

We can determine the crossover scaling function by integrating this equation.  Taking $T_{0}=T_{0}^{0}$ at $l=0$,

\begin{equation}
\frac{T_{0}^{0}(1-T_{0}^{0})^{\frac{1}{\frac{2}{\gamma}-1}}}{\left(1-\frac{2T_{0}^{0}}{\gamma}\right)^{\frac{2}{\gamma}\left(\frac{1}{\frac{2}{\gamma}-1}\right)}}e^{-2\gamma(\epsilon_{+} + \epsilon_{-})^{2}l} = \frac{T_{0}(1-T_{0})^{\frac{1}{\frac{2}{\gamma}-1}}}{\left(1-\frac{2T_{0}}{\gamma}\right)^{\frac{2}{\gamma}\left(\frac{1}{\frac{2}{\gamma}-1}\right)}}.
\end{equation}

\begin{figure}
\centering
\scalebox{0.41}{\includegraphics*{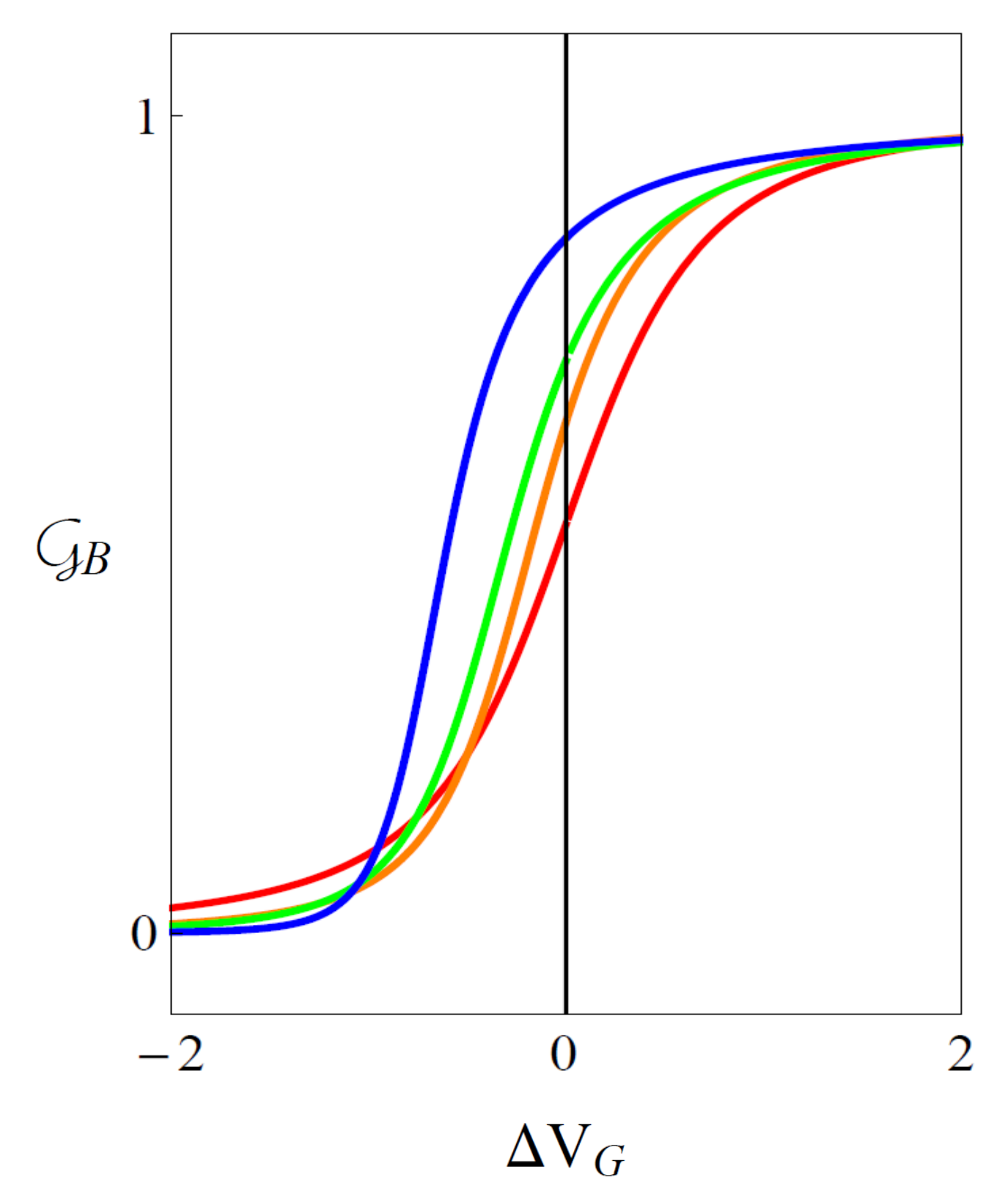}}
\caption{The universal scaling function $\mathcal{G}_{B}$ which characterizes the II-M phase transition, plotted for $T/c^{1/\alpha}=10^{-3}$ $V^{1/\alpha}$ and $g_{+} = 1.212$.  The blue, green, orange, and red curves are plotted at $g_{-}=1.019$, $1.047$, $1.071$, and $1.212$ respectively.  The critical value of $\mathcal{G}_{B}$ for which the conductance flow changes from the II phase to the M phase occurs at the intersection of each curve with the $\Delta V_{G}=0$ line, and varies as a function of the ratio of the interaction strengths $g_{-}/g_{+}$.  When $g_{-}=g_{+}$, $\mathcal{G}_{B}$ takes on the same functional form as $\mathcal{G}_{A}$ in Fig.~\ref{FigTemp}, though with a different critical exponent $\alpha_{B}\neq\alpha_{A}$.  At $T=0$ K, all of these curves collapse onto the same step function; they are increasingly distinguishable as temperature is increased.}
\label{FigScaling}
\end{figure}

As before, we can cut off the renormalization group flow at finite temperature $T\propto\Lambda e^{-l}$ and note that the gate voltage $V_{G}=V_{G,B}^{*}$ when $T_{0}^{0}$ passes through $\gamma/2$, such that near the transition $\Delta V_{G} \propto T_{0}^{0} - \gamma/2$.  Again taking $\Delta V_{G},T\rightarrow 0$ at an arbitrary ratio, we can finally arrive at an implicit equation for $\mathcal{G}_{B}(X)$,

\begin{equation}
\frac{1}{\left(\frac{2X}{\gamma}\right)^{2}}=\frac{\mathcal{G}_{B}^{(2-\gamma)}(1-\mathcal{G}_{B})^{\gamma}}{\left(1-\frac{2\mathcal{G}_{B}}{\gamma}\right)^{2}}
\label{GBscaling}
\end{equation}   

where $T_{0} = \mathcal{G}_{B}$ such that $X\propto\Delta V_{G} /T^{(\epsilon_{+}+\epsilon_{-})^{2}\gamma\left(2-\gamma\right)}$.  We can note that this equation reduces to Eq.~(\ref{GAscaling}) for $\gamma=1$.  That case represents $\epsilon_{-}\rightarrow\epsilon_{+}$, for which $T_{0}$ and $T_{\pi}$ act as independent copies of the Quantum Spin Hall problem in Ref.~\onlinecite{teokane}, but in a higher-dimensional space.  Therefore,

\begin{eqnarray}
\alpha_{B} &=& (\epsilon_{+}+\epsilon_{-})^{2}\gamma\left(2-\gamma\right) \nonumber \\
&=& \alpha_{A}\left(2-\gamma\right)
\end{eqnarray}

is the universal critical exponent for the II-M transition. Confirming the relationship to the QSH point contact problem, setting $\gamma=1$ gives $\left. \alpha_{B}\right|_{\epsilon_{-}=\epsilon_{+}}=\left. 2\alpha_{A}\right|_{\epsilon_{-}=0}=8\alpha_{QSH}$.  Figure~\ref{FigTemp} shows $\mathcal{G}_{B}$ as a function of $\Delta V_{G}$ at different temperatures, noting that at zero temperature it is also a step function, indistinguishable from $\mathcal{G}_{A}$, but that at finite temperature it is defined by a crossover value of $\gamma/2$ which in general differs from that of $\mathcal{G}_{A}$ (Fig.~\ref{FigScaling}). 

Taking advantage of the duality between the II and CC phases, we can relate the remaining conductance crossover function, one which characterizes the M-CC phase transition, to $\mathcal{G}_{B}$.  We can write the conductance in its scaling form

\begin{equation}
G_{XX,C} = \frac{4e^{2}}{h}+\frac{4e^{2}}{h}\mathcal{G}_{C}\left(c\frac{\Delta V_{G}}{T^{\alpha_{C}}}\right)
\end{equation}

where $c$ is yet another non-universal constant and $C$ denotes that M-CC transition.  By pinch-off symmetry, we know that $G_{YY}$ for the CC$\rightarrow$M transition has to be equivalent to $G_{XX}$ for the II$\rightarrow$M transition, therefore for the M$\rightarrow$CC transition, 

\begin{equation}
G_{YY,C} = \frac{4e^{2}}{h}\mathcal{G}_{B}\left(-c\frac{\Delta V_{G}}{T^{\alpha_{B}}}\right)
\end{equation}

and utilizing Eq.~(\ref{2term}), 

\begin{equation}
G_{XX,C} = \frac{8e^{2}}{h}-\frac{4e^{2}}{h}\mathcal{G}_{B}(-X) 
\end{equation}

such that we finally deduce

\begin{equation}
\mathcal{G}_{C}(X) = 1 - \mathcal{G}_{B}(-X)
\end{equation}

where $X\propto\Delta V_{G}/T^{\alpha_{C}}$ and 

\begin{equation}
\alpha_{C}=\alpha_{B}
\end{equation}

for weak interactions.  

\section{Discussion}

In this paper, we have computed the conductance signatures of the four-terminal intersection of two bilayer graphene domain walls.  These domain walls can be induced by the presence of a perpendicular electric field and a change in either electric field direction or interlayer stacking.  When valley-index is conserved, the domain walls are Luttinger liquids described with two non-trivial interaction parameters $g_{\pm}$.  The junction is analogous to a point contact and can be analyzed naturally using the language of quantum point contacts.  As with a Quantum Spin Hall point contact, the physics of the junction is best understood in terms of reduced, two-terminal conductances.  When interactions are strongly attractive ($g_{\pm}< 1/4$) or strongly repulsive ($g_{\pm}> 4$), the left-to-right conductance can be strictly dominated by nonzero charge and valley conductances respectively.  For weaker interactions ($g_{\pm}\approx 1$), both left-to-right conductances are nonzero and there exist several stable phases characterized by single-electron tunneling.  Transitions between these phases are governed at low temperatures by universal scaling functions and critical exponents, which differ from those in the single-band QSH case and are functions of the two Luttinger parameters.  

We now briefly discuss the task of experimentally measuring the physics in this paper.  First and foremost, the existence of a single domain wall in bilayer graphene requires prohibiting scattering between valleys $K$ and $K'$.  Valley-index-breaking perturbations are strongly relevant and will significantly change the physics in both isolated domain walls and for the junction structures at their intersections.  To this end, short-range disorder must be kept smooth on the scale of the lattice.  Recent experiments have shown promising results in this direction, with fabricated samples showing single domain wall conductances up to nearly the quantized clean limit of $4e^{2}/h$~\cite{natureEx}.  Beyond these results, one should then attempt to verify the Luttinger liquid physics at a single domain wall by measuring the tunneling conductance at several low temperatures and its collapse onto a universal scaling function with critical exponent $\alpha_{T}$.  In junction structures, the conservation of valley index can be confirmed by measuring the quantization of the reduced conductance element $G_{ZZ}$.  

Creating a multi-terminal junction like the ones we describe poses several fabrication and analysis difficulties which must be overcome to measure the point-contact physics in this paper.  Forming a four-terminal junction of electric-field-induced domain walls requires patterning leads on the top and bottom of each of the four bulk regions, as well as a gate on the junction to control the weak-interaction pinch-off transition.  Forming a junction from layer-stacking domain walls requires patterning fewer leads, but as clearly demonstrated in the samples in Ref.~\onlinecite{Alden}, the three-fold symmetry of the underlying graphene lattice restricts intersections of these domain walls to be six-terminal structures.  Conductance transitions in these six-terminal structures can be calculated and analyzed using the framework established in this paper, though the task will be algebraically more intensive.  

Tuning the two Luttinger parameters $g_{\pm}$ can be accomplished through turning a combination of experimental knobs.  The strength of the overall effective Coulomb interaction can be altered for the domain wall states by tuning their widths with the strength of the perpendicular electric field.  These changes will be mainly reflected in $g_{+}$, as it contains long-range contributions from the overall Coulomb interaction.  The other Luttinger parameter, however, can only be adjusted by tuning short-range interactions.  This can be accomplished by testing samples on a variety of substrates.  Working in order of increasing dielectric strength, one can work through a suspended sample, a silicon dioxide substrate, or a boron nitride substrate to tune down $g_{-}$.  

Several simplifications and assumptions in this paper may not be exactly present under experimental conditions.  The assumption that all domain walls in the sample have the same values of the Luttinger parameters requires that the perpendicular electric field strength be globally uniform in magnitude in the bulk regions and change in direction similarly smoothly across electric-field-induced domain walls.  It also requires that there are no strong local variations in the dielectric strength and coupling of the underlying substrate.  The physics of junctions may be significantly altered if these conditions are not realized experimentally; we have not investigated point contacts at the intersections of two domain walls with differing interaction strengths.  We also only calculated universal scaling functions to leading order in the interactions about the noninteracting point $g_{+}=g_{-}=1$.  When interactions become stronger and single-electron tunneling is no longer the least irrelevant operator, the critical exponents $\alpha_{A/B}$ in the left-to-right conductance transitions may change significantly in their dependences on the interaction strengths, as they do in Ref.~\onlinecite{teokane}.  Finally, we assumed that the Fermi energy was exactly at the particle-hole-symmetric point such that $v_{F,0}=v_{F,\pi}=v_{F}$.  In practice, it will be quite difficult to exactly tune the Fermi energy to this point, and so for most experimental realizations, $0\leftrightarrow\pi$ exchange symmetry in the variables will be broken.  For our analysis, the effects of this can be realized by replacing the equal band-index-exchange symmetry which we exploited with one which flips band index and rescales variables by $v_{F,0}/v_{F,\pi}$.  This will result in changes to the scaling dimension calculations in Section~\ref{sec:fourterminal} and a relaxation of mirror symmetry about $T_{0}=T_{\pi}$ in Figure~\ref{FigRG}.  

\acknowledgments

This work was supported by a Simons Investigator award from the Simons Foundation to Charles Kane.  Fan Zhang was supported by UT Dallas research enhancement funds. 

\begin{appendix}

\section{Stability of Quantum Critical Points on the $\theta_{i}=0$ Surface}
\label{appendix:stability}

In this appendix, we confirm analytically the stability of the two classes of quantum critical points in Fig.~\ref{FigRG} which control the transitions between single-electron-tunneling phases.  Here, we begin with the general flow equation for elements of the S-matrix (Eq.~(\ref{Sflow})), leaving free the $\theta_{i}$ and $\phi_{ij}$ variables in the full S-matrix formulation from~\ref{sec:noninteracting}.  As the complexity of this calculation greatly grows with each power of $\epsilon_{-}$ taken into consideration, we will only here carry out our stability analysis to quadratic order in $\epsilon_{+}$ and linear order in $\epsilon_{-}$.  Our methodology can be used to analytically calculate higher-order terms, but we believe it reasonable to truncate the calculation at this order and that the stability should carry over to the $\mathcal{O}(\epsilon_{-}^{2})$ calculation used to produce the phase diagrams and scaling functions in~\ref{sec:interactingRG}.  Additionally, numerical estimates find $g_{-}<g_{+}$ for a large range of system parameters~\cite{affleck}, therefore the limitation to terms of order $\mathcal{O}(\epsilon_{+}\epsilon_{-},\epsilon_{+}^{2})$ is physically motivated.  

We can exploit the matrix structure of the diagrammatics by taking several traces of S-matrix products.  Derivatives of these traces exploit the matrix structure of the flow equation, effectively closing the external legs of the diagrams in Fig.~\ref{FigDiagram} into additional loops.  We will follow through this calculation completely for a single trace ($\Tr[S_{ij}^{\dagger}\sigma_{z}S_{ij}]$) and then show how taking linear combinations of these traces results in a flow equation for $\theta_{1}$.  All remaining $d\theta_{i}/dl$ can be obtained by exploiting cyclic reindexing and pinch-off symmetries.  

\subsection{Flow Equations for $\theta_{i}$ Band Mixing Angles and $\phi_{ij}$ Scattering Phases}
\label{appendix:thetaphi}

To begin this process, let us rewrite S-matrix elements in the $2\times 2$ band space using a simplified version of Eqs.~(\ref{Smatrix}),~(\ref{Umatrix}):

\begin{eqnarray}
S^{\alpha\beta}_{ij} &=& (U_{i}^{\dagger}D_{ij}U_{j})^{\alpha\beta} \nonumber \\
U_{i} &=& e^{i\frac{\theta_{i}}{2}\sigma_{y}} \nonumber \\
D_{ij} &=& a_{ij}\mathbb{1} + b_{ij}\sigma_{z} \nonumber \\
a_{ij} &=& \frac{1}{2}\left(d_{ij}^{0} + d_{ij}^{\pi}\right) \nonumber \\
b_{ij} &=& \frac{1}{2}\left(d_{ij}^{0} - d_{ij}^{\pi}\right)
\end{eqnarray}

\begin{equation}
d^{a} = \left(\begin{array}{cccc}
0 & t^{a}e^{i\phi_{A}^{a}} & 0 & r^{a}e^{i\phi_{B}^{a}} \\
t^{a}e^{i\phi_{A}^{a}} & 0 & -r^{a}e^{i\phi_{C}^{a}} & 0 \\
0 & -r^{a}e^{i\phi_{C}^{a}} & 0 & t^{a}e^{i\phi_{D}^{a}} \\
-r^{a}e^{i\phi_{C}^{a}} & 0 & t^{a}e^{i\phi_{D}^{a}} & 0
\end{array}\right)
\end{equation}

where $d_{ij}^{a}$ is a collection of scalars in the band index space with $a=0,\pi$, $\phi^{0}_{ij}=-\phi^{\pi}_{ij}=\phi_{ij}$ and all matrix operations only involve the $\alpha,\beta=0,\pi$ indexes.  Transmission probabilities are normalized for each band such that $(t^{a})^{2} + (r^{a})^{2} = 1$ and the labeling $A-C$ on the phases corresponds to:

\begin{eqnarray}
\phi_{A} &=& \phi_{21} = \phi_{2}-\phi_{1} \nonumber \\
\phi_{B} &=& \phi_{41} \nonumber \\
\phi_{C} &=& \phi_{23} \nonumber \\
\phi_{D} &=& \phi_{43} = \phi_{B} - \phi_{A} + \phi_{C}. 
\label{phidef}
\end{eqnarray}

We'll begin by calculating three traces:

\begin{eqnarray}
\Tr[S_{ij}^{\dagger}S_{ij}] &=& |d_{ij}^{0}|^{2} + |d_{ij}^{\pi}|^{2} \nonumber \\
\Tr[S_{ij}^{\dagger}\sigma_{z}S_{ij}] &=& \left(|d_{ij}^{0}|^{2} - |d_{ij}^{\pi}|^{2}\right)\cos{\theta_{i}} \nonumber \\
\Tr[S_{ij}^{\dagger}\sigma_{x}S_{ij}] &=& \left(|d_{ij}^{0}|^{2} - |d_{ij}^{\pi}|^{2}\right)\sin{\theta_{i}}.
\label{2S}
\end{eqnarray}

We can then differentiate and take weighted linear combinations of these traces to obtain explicit flow equations for $\theta_{i}$.  Specializing to $i,j=1,2$, 

\begin{eqnarray}
\left(T_{0} - T_{\pi}\right)\frac{d\theta_{1}}{dl} &=& (\cos{\theta_{1}})\frac{d}{dl}\Tr[S_{12}^{\dagger}\sigma_{x}S_{12}] \nonumber \\
&-& (\sin{\theta_{1}})\frac{d}{dl}\Tr[S_{12}^{\dagger}\sigma_{z}S_{12}] 
\end{eqnarray}

where $T_{0/\pi} = |t^{0/\pi}|^{2}$.  Now, we can examine, in detail, the process of using the diagrammatics (Eq.~(\ref{Sflow})) to calculate one of these derivatives, namely $\frac{d}{dl}\Tr[S_{12}^{\dagger}\sigma_{z}S_{12}]$.  All other trace derivatives, while varying in signs and specifics, follow procedurally from this example.  

First, by using the cyclic index definition of the trace, we can see that the derivative of the trace is equal to the trace of the chain rule derivative:

\begin{eqnarray}
\frac{d}{dl}\Tr[S_{12}^{\dagger}\sigma_{z}S_{12}] &=& \Tr[\frac{dS^{\dagger}_{12}}{dl}\sigma_{z}S_{12}] + \Tr[S^{\dagger}_{12}\sigma_{z}\frac{dS_{12}}{dl}] \nonumber \\
&=& \Tr[S^{\dagger}_{12}\sigma_{z}\frac{dS_{12}}{dl}] + C.C.
\end{eqnarray}

where for the second equality we exploited the cyclic nature of the trace to reduce this step to the calculation of one, albeit large, trace.  In this case, the trace of the derivative is real, but in the next section where we calculate $d\phi_{ij}/dl$, it will not be and the addition of the complex conjugate cannot be overlooked.   

From here, the calculation amounts to taking the traces of terms which contain products of two or four S-matrices.  While the products of two, in the form of Eq.~(\ref{2S}), can be calculated by rote algebra without much difficulty, the terms with four S-matrices require a bit more manipulation.

We calculate the traces of products of four S-matrices by both recognizing a pattern in the assignment of signs to the products of $a_{ij},b_{ij}$ and with a careful treatment of commutivity issues.  Consider first the simplest example, $\Tr[S^{\dagger}_{l2}S_{lk}S^{\dagger}_{1k}S_{12}]$.  Without any additional Pauli matrices, the $U_{i}$ rotations all cancel out pairwise, resulting in:

\begin{eqnarray}
\Tr[S^{\dagger}_{l2}S_{lk}S^{\dagger}_{1k}S_{12}] = \Tr[(a_{l2}^{*}\mathbb{1} + b_{l2}^{*}\sigma_{z})(a_{lk}\mathbb{1}+b_{lk}\sigma_{z}) \nonumber \\
(a_{1k}^{*}\mathbb{1} + b_{1k}^{*}\sigma_{z})(a_{12}\mathbb{1} + b_{12}\sigma_{z})].\ 
\end{eqnarray}

One might be concerned that converting this to a useful form, one with $d^{0/\pi}_{ij}$ where S-matrix elements can just be read off, would be a daunting and terrible task.  However, converting to $d^{0/\pi}_{ij}$ basis is equivalent to summing over all of the ways to choose $+$ and $-$ signs for the cross terms, and so for every single one of these four S-matrix traces, \emph{all of the cross terms cancel}.  Our trace is reduced to the quite simple form:

\begin{equation}
\Tr[S^{\dagger}_{l2}S_{lk}S^{\dagger}_{1k}S_{12}]=\sum_{a=0,\pi}(d_{l2}^{a})^{*}d_{lk}^{a}(d_{1k}^{a})^{*}d_{12}^{a}.  
\label{simple4trace}
\end{equation}

Worth noting is that this picture is significantly complicated by the addition of Pauli matrices between S-matrix factors, due to the fact that $[U_{i},\sigma_{z/x}]\neq 0$.  While the complexity doesn't significantly increase for the addition of a single Pauli matrix, as it can be absorbed into the definition of $D_{ij}$, it does for two or more Pauli matrices.  This can be seen at the level of the two-S-matrix trace, where commutivity issues lead to the addition of a second term:

\begin{eqnarray}
\Tr[\sigma_{z}S^{\dagger}_{ij}\sigma_{z}S_{ij}]&=&\cos{\theta_{i}}\cos{\theta_{j}}\left[|d_{ij}^{0}|^{2} + |d_{ij}^{\pi}|^{2}\right] \nonumber \\
&+& \sin{\theta_{i}}\sin{\theta_{j}}\left[d_{ij}^{0}(d_{ij}^{\pi})^{*}+(d_{ij}^{0})^{*}d_{ij}^{\pi}\right].\nonumber \\  
\end{eqnarray}

For the four-S-matrix traces, the weighting of the cross terms is altered by the anticommutivity of the Pauli matrices and while only two terms remain for each trigonometric function of $\theta_{i}$, they are different than the simple form of Eq.~(\ref{simple4trace}) and contain possible terms which mix $0$ and $\pi$.

As the number of Pauli matrices inserted into these traces increases linearly with the power of $\epsilon_{-}$ to which we expand, we have chosen for the sake of simplicity and clarity to expand only to linear order in $\epsilon_{-}$.  Though our analysis in section~\ref{section:qcps} continues to $\mathcal{O}(\epsilon_{-}^{2})$, we believe that the stability deduced here should carry over to higher order terms.  

With our calculation machinery established, we can produce a flow equation for the $0\leftrightarrow\pi$ mixing at lead $1$:

\begin{eqnarray}
\frac{d\theta_{1}}{dl} =&\ & -\epsilon_{+}\epsilon_{-}\bigg\{\sin{2\theta_{1}}[T_{0}(1-T_{\pi}) + T_{\pi}(1-T_{0})] \nonumber \\
&+& \sin{2\theta_{2}}\cos\left(2\phi_{21}\right)\sqrt{T_{0}T_{\pi}}[2-T_{0}-T_{\pi}] \nonumber \\
&+& 2\sin{2\theta_{3}}\cos\left(2\phi_{31}\right)\sqrt{T_{0}(1-T_{0})T_{\pi}(1-T_{\pi})} \nonumber \\
&+&\sin{2\theta_{4}}\cos\left(2\phi_{41}\right)\sqrt{(1-T_{0})(1-T_{\pi})}[T_{0}+T_{\pi}]\bigg\}.\nonumber \\
\end{eqnarray}

Flow equations for the remaining three mixing angles can be generated by exploiting underlying symmetries of the problem.  The set of nine independent variables which characterizes the S-matrix obeys three symmetries.  Two ``pinch-off'' symmetries exist; the duality between the fully closed (II) and fully open (CC) single-electron phases implies that the system of flow equations is invariant with respect to the exchange $T_{0/\pi}\leftrightarrow(1-T_{0/\pi})$ and \emph{either} the exchange of lead indexes $2\leftrightarrow 4$ \emph{or} $1\leftrightarrow 3$.  The third symmetry is a cyclic relabeling of the lead indexes as well as an exchange of the definitions of pinched off and open, due to the system's invariance under properly-treated (with respect to valley) $90^{\circ}$ rotations.  Therefore the system is also invariant under the exchange $T_{0/\pi}\leftrightarrow(1-T_{0/\pi})$ and $1\rightarrow 2$, $2\rightarrow 3$,$3\rightarrow 4$, and $4\rightarrow 1$.  Independent calculations of $d\theta_{i}/dl$ confirm these properties.  

We can see immediately that for $\epsilon_{-}=0$, all $\theta_{i}$ are marginal.  In this case, the system has U(2) symmetry at each lead and all band index rotations can be gauged out.  The dependence of $d\theta_{i}/dl$ on $\theta_{j\neq i}$ can also be suppressed by tuning the scattering phase $\phi_{ij}$ closer to $\pi/4$, which amounts to having a $\pi/2$ scattering phase difference between the $0$ and $\pi$ bands.    

Calculating the flow equations for scattering phases $\phi_{ij}$ requires, conversely, tracing over open paths in diagrams, which allows phase to accumulate throughout the summation instead of being canceled out pairwise as frequently occurred in the calculation of $d\theta_{i}/dl$.  Specializing for the moment towards obtaining $d\phi_{A}/dl$, we can note the following:

\begin{eqnarray}
\Tr[S_{12}] &=& (t_{0}e^{i\phi_{A}} + t_{\pi}e^{-i\phi_{A}})\cos\left(\frac{\theta_{1}-\theta_{2}}{2}\right) \nonumber \\
|\Tr[S_{12}]|^{2} &=& \cos^{2}\left(\frac{\theta_{1}-\theta_{2}}{2}\right)\big[T_{0} + T_{\pi} + 2\sqrt{T_{0}T_{\pi}}\cos{2\phi_{A}}\big] \nonumber \\
\end{eqnarray}

where the correspondence between the subscripts $A-D$ and the lead indexes $i,j$ comes from Eq.~(\ref{phidef}).  This can be differentiated, and, with a considerable amount of algebra, used to obtain first order flow equations for the scattering phases.  While the specifics of this calculation differ from those of obtaining the $d\theta_{i}/dl$, the key point about the summation over $+$ and $-$ possibilities when converting to the $d^{0/\pi}_{ij}$ basis remains for both the one- and three-S-matrix products here, again greatly simplifying the algebra for the required trace calculations.  Utilizing this fact, we obtained an explicit flow equation for scattering phase $\phi_{A}$:

\begin{widetext}
\begin{eqnarray}
\frac{d\phi_{A}}{dl} = \frac{1}{2}&\epsilon_{+}\epsilon_{-}&\tan\left(\frac{\theta_{1}-\theta_{2}}{2}\right)\bigg\{(\sin{2\theta_{1}}-\sin{2\theta_{2}})\sin\left(2\phi_{A}\right)\sqrt{T_{0}T_{\pi}}(2-T_{0}-T_{\pi}) \nonumber \\
&+& \sin{2\theta_{3}}\sqrt{(1-T_{0})(1-T_{\pi})}\left[(T_{0}+T_{\pi})\sin\left(2\phi_{C}\right) - 2\sqrt{T_{0}T_{\pi}}\sin\left(2(\phi_{A}-\phi_{C})\right)\right] \nonumber \\
&-&\sin{2\theta_{4}}\sqrt{(1-T_{0})(1-T{\pi})}\left[(T_{0}+T_{\pi})\sin\left(2\phi_{B}\right) - 2\sqrt{T_{0}T_{\pi}}\sin\left(2(\phi_{A}-\phi_{B})\right)\right]\bigg\}.
\end{eqnarray}
\end{widetext}

The flow equations for $\phi_{B-D}$ can be obtained by exploiting pinch-off and cyclic symmetries as well as the redundancy of $\phi_{D}$ (Eq.~(\ref{phidef})).  As with the $\theta_{i}$, one can observe that for the $U(2)$ symmetric case of $\epsilon_{-}=0$, all $\phi_{ij}$ are also marginal and can be gauged away.  One can also observe that $\frac{d\phi_{ij}}{dl}=0$ when $\theta_{i}=\theta_{j}$, as in that case there is no relative interband scattering between leads $i$ and $j$ and $\phi_{ij}$ can be gauged out.  Only two unique stability calculations are required, as the four critical points on the boundary of the square are related by pinch-off and band-index-exchange symmetries.    

\subsection{Quantum Critical Point Stability}
\label{appendix:QCPstab}

We can now utilize the flow equations for $\theta_{i}$ and, to a lesser extent, $\phi_{ij}$ to determine the stability of the $\theta_{i}=\phi_{ij}=0$ surface quantum critical points in Fig.~\ref{FigRG}.

We will begin by considering the $T_{0}=T_{\pi}=1/2$ central quantum critical point, which controls the direct transition between the fully pinched-off (II) and fully-open (CC) phases.  Expanding the $\theta_{i}$ to linear order:

\begin{equation}
\frac{d}{dl}\left(\begin{array}{c}
\theta_{1} \\
\theta_{2} \\
\theta_{3} \\
\theta_{4} 
\end{array}\right) = -\epsilon_{-}\epsilon_{+}\mathbb{M}\left(\begin{array}{c}
\theta_{1} \\
\theta_{2} \\
\theta_{3} \\
\theta_{4} 
\end{array}\right)
\end{equation}

\begin{equation}
\mathbb{M}_{ij}=\cos{2\phi_{ij}}
\end{equation}

which gives two Lyapunov exponents of zero and two of $\lambda_{\pm}=-2\epsilon_{+}\epsilon_{-}\left\{1\pm\frac{1}{4}\sqrt{\sum_{i,j}\cos{4\phi_{ij}}}\right\}$.  Both $\lambda_{\pm}$ eigenvalues are either zero or negative for all choices of the independent $\phi_{A-C}$.  Given the high symmetry of this central quantum critical point, it is unlikely that the $\lambda=0$ marginal directions correspond to instabilities at higher orders.  

The four external critical points which control the transitions to and from the mixed (M) phase can be handled similarly, but with an attention to the expansions used to arrive at this analysis.  As our model is invariant under global exchange $0\leftrightarrow\pi$ and the pinch-off symmetries, we perform stability analysis on just one of the four critical points and relate the rest by symmetry.  

Choosing the critical point at $T_{0}=0, T_{\pi}=\gamma/2$, we can again construct a stability matrix: 

\begin{equation}
\frac{d}{dl}\left(\begin{array}{c}
\theta_{1} \\
\theta_{2} \\
\theta_{3} \\
\theta_{4} 
\end{array}\right) = -\epsilon_{+}\epsilon_{-}\gamma\mathbb{N}\left(\begin{array}{c}
\theta_{1} \\
\theta_{2} \\
\theta_{3} \\
\theta_{4} 
\end{array}\right)
\end{equation}

\begin{equation}
\mathbb{N}=\mathbb{1} + \sqrt{1-\frac{\gamma}{2}}\left(\begin{array}{cccc}
0 & 0 & 0 & \cos{2\phi_{B}} \\
0 & 0 & \cos{2\phi_{C}} & 0 \\
0 & \cos{2\phi_{C}} & 0 \\
\cos{2\phi_{B}} & 0 & 0 
\end{array}\right). 
\end{equation}

Acknowledging that, as we have expanded to linear order in $\epsilon_{-}$, $\gamma\approx2-4\epsilon_{-}/\epsilon{+}$, all of the off-diagonal terms become infinitesimal and we are left with:

\begin{equation}
\frac{d}{dl}\left(\begin{array}{c}
\theta_{1} \\
\theta_{2} \\
\theta_{3} \\
\theta_{4} 
\end{array}\right) = -2\epsilon_{+}\epsilon_{-}\left(1-\frac{2\epsilon_{-}}{\epsilon_{+}}\right)\left(\begin{array}{c}
\theta_{1} \\
\theta_{2} \\
\theta_{3} \\
\theta_{4} 
\end{array}\right)
\end{equation}

which clearly has strictly negative Lyapunov exponents up to this order of expansion.

\section{Resolving the Fixed Line in the U(2) Symmetric Case}
\label{appendix:bosonization}

The flow diagram in Fig.~\ref{FigQuad}, obtained by perturbation theory to quadratic order in the interactions, suitably describes the transport physics of single-electron-tunneling phases for most ranges of infinitesimal interactions $\epsilon_{+}$ and $\epsilon_{-}$.  However, when $\epsilon_{-}=0$, Eqs.~(\ref{flowEQ}) become symmetric under arbitrary U(2) transformations in band index space and describe a fixed \emph{line} $T_{0} + T_{\pi} = 1$.  In this appendix, we use bosonization at the $T_{0}=1$, $T_{\pi}=0$ ``M-phase'' fixed point to determine whether this fixed line is a genuine physical phenomenon or an artifact of our $\mathcal{O}(\epsilon_{\pm}^{2})$ perturbation theory.  The first section of this appendix details the derivation of a Euclidean action about an M fixed point, working from the action for the fully-open CC phase.  The second section utilizes that action to calculate flow equations beyond quadratic order in the vicinity of the M phase, resolving the fixed line to additional fixed points for weak-to-moderate interactions.

\subsection{Deriving the Euclidean Action for the $T_{0}=1, T_{\pi}=0$ M-Phase Fixed Point}
\label{appendix:action}

To examine the higher-order behavior of tunneling processes about the $T_{0}=1,\ T_{\pi}=0$ corner fixed point in Eq.~(\ref{flowEQ}), we have to first obtain a theory for that fixed point in terms of our existing theory for the $T_{0,\pi}=1$, fully-open fixed point.  As we will be focusing on the fate of the quadratic-order fixed line at $g_{-}=1$, we will for the purposes of this calculation specialize to $g_{+}=g$,\ $g_{-}=1$.  Integrating out the $\phi_{+/-,c/s,\rho/v}$ sectors in Eq.~(\ref{spinfulLL}), we can propose as a starting point the Euclidean action about the CC fixed point:

\begin{equation}
S_{CC} = \frac{1}{4\pi\beta}\sum_{\alpha=c,s}\sum_{\omega_{n}}|\omega_{n}|\vec{\theta}_{\alpha}^{vT}\mathbb{g}_{\alpha}\vec{\theta}_{\alpha}^{v}
\end{equation}

where

\begin{equation}
\vec{\theta}_{\alpha}^{v}=\left(\begin{array}{c}
\theta_{+\alpha\rho} \\
\theta_{+\alpha v} \\
\theta_{-\alpha\rho} \\
\theta_{-\alpha v}
\end{array}\right),
\end{equation}

\begin{equation}
\mathbb{g}_{c} = \left(\begin{array}{cccc}
g & 0 & 0 & 0 \\
0 & \frac{1}{g} & 0 & 0 \\
0 & 0 & 1 & 0 \\
0 & 0 & 0 & 1
\end{array}\right),\ \mathbb{g}_{s} = \left(\begin{array}{cccc}
1 & 0 & 0 & 0 \\
0 & 1 & 0 & 0 \\
0 & 0 & 1 & 0 \\
0 & 0 & 0 & 1
\end{array}\right).
\end{equation}

We can introduce weak tunneling processes, 

\begin{eqnarray}
V_{0/\pi} &=& \frac{v_{0/\pi}}{2}\left[\psi_{0/\pi,\uparrow RK}^{\dagger}\psi_{0/\pi,\uparrow LK} + K\leftrightarrow K' + \uparrow\leftrightarrow\downarrow +\ H.C.\right] \nonumber \\
&=& v_{0/\pi}\sum_{\sigma,u}\cos\left(\theta_{0/\pi,\sigma u}\right)
\end{eqnarray}

where $\sigma=\uparrow,\downarrow$ and $u = K,K'$.  

These processes will generally be present within the vicinity of the $T_{0}=T_{\pi}=1$ fixed point.  Both of these processes are just single-electron tunneling, so as described in~\ref{sec:fourterminal}, they are marginal or irrelevant ($\Delta(v_0)=\Delta(v_\pi)=\frac{1}{8}[g + 1/g + 6]$) for all values of $g$, such that we may arbitrarily increase the coupling strength $v_{0/\pi}$ without new, additional tunneling processes becoming relevant.  Respecting time-reversal, valley-index, and band-index symmetries, $V_{0/\pi}$ are restricted to live in the plane of Fig.~\ref{FigRG} and therefore turning up $v_{0/\pi}$ represents motion away from the CC fixed point along the $T_{0/\pi}$ axes respectively.  

For this reason, we need to return $S_{CC}$ to the $0,\pi$ basis so that we can generate an M phase action by taking the tunneling strength $v_{\pi}\rightarrow\infty$ and ``pinch off'' just the $\pi$ band.  Additionally, we will find it easier to work in a basis for which the tunneling operators $V_{0/\pi}$ are just sums of cosines.  From our bosonization work in~\ref{sec:model} and utilized earlier in this appendix, we recall that $\theta_{\pm\sigma u}=\theta_{0\sigma u}\pm\theta_{\pi\sigma u}$.  We will also need to transform valley indexes utilizing the property that $\theta_{a\sigma, K/K'} = \theta_{a\sigma\rho}\pm\theta_{a\sigma v}$ where $a=0,\pi$.  Combining these definitions, we create the unitary change-of-basis 

\begin{equation}
\vec{\theta}^{v}_{\alpha}=\left(\begin{array}{c}
\theta_{+\alpha\rho} \\
\theta_{+\alpha v} \\
\theta_{-\alpha\rho} \\
\theta_{-\alpha v}
\end{array}\right)=\mathbb{Y}\left(\begin{array}{c}
\theta_{0\alpha K} \\
\theta_{0\alpha K'} \\
\theta_{\pi\alpha K} \\
\theta_{\pi\alpha K'}
\end{array}\right) = \mathbb{Y}\vec{\theta}_{\alpha}
\end{equation}

where $\alpha=c,s$ and 

\begin{equation}
\mathbb{Y} = \frac{1}{2}\left(\begin{array}{cccc}
1 & 1 & 1 & 1 \\
1 & -1 & 1 & -1 \\
1 & 1 & -1 & -1 \\
1 & -1 & -1 & 1
\end{array}\right)  
\end{equation}

and we can note that $\mathbb{Y}^{T}=\mathbb{Y}^{-1}=\mathbb{Y}$.  Under this transformation, the action becomes

\begin{equation}
S_{CC} = \frac{1}{4\pi\beta}\sum_{\alpha=c,s}\sum_{\omega_{n}}|\omega_{n}|\vec{\theta}_{\alpha}^{T}\mathbb{Y}\mathbb{g}_{\alpha}\mathbb{Y}\vec{\theta}_{\alpha}
\end{equation}

where $\mathbb{Y}\mathbb{g}_{c}\mathbb{Y}$ separates into blocks:

\begin{equation}
\mathbb{Y}\mathbb{g}_{c}\mathbb{Y} = \frac{1}{4}\left(\begin{array}{cc}
\mathbb{A} & \mathbb{B} \\
\mathbb{B} & \mathbb{A}
\end{array}\right)
\end{equation}

where

\begin{eqnarray}
\mathbb{A} &=& \left(\begin{array}{cc}
g+ \frac{1}{g} + 2 & g- \frac{1}{g} \\
g-\frac{1}{g}& g+ \frac{1}{g} + 2
\end{array}\right) \nonumber \\
&=& \left(g+1/g + 2\right)\mathbb{I} + \left(g-1/g\right)\sigma^{x}\nonumber \\ \nonumber \\
\mathbb{B} &=& \left(\begin{array}{cc}
g+ \frac{1}{g} - 2 & g- \frac{1}{g} \\
g-\frac{1}{g}& g+ \frac{1}{g} - 2
\end{array}\right) \nonumber \\ 
&=&\left(g+1/g - 2\right)\mathbb{I} + \left(g-1/g\right)\sigma^{x}=\mathbb{A}-4\mathbb{I}.  
\end{eqnarray}

As the spin sector is noninteracting, $\mathbb{Y}\mathbb{g}_{s}\mathbb{Y}=\mathbb{I}$.

To complete the transformation into the basis of 

\begin{equation}
V_{CC} = V_{0} + V_{\pi},
\end{equation}

we now need to address the spin degree of freedom.  Utilizing the definitions of the spin and charge sectors (Eqs.~(\ref{Defs})),

\begin{equation}
\left(\begin{array}{c}
\vec{\theta}_{c} \\
\vec{\theta}_{s} 
\end{array}\right)=\left(\begin{array}{cc}
\mathbb{I} & \mathbb{I} \\
\mathbb{I} & -\mathbb{I}
\end{array}\right)\left(\begin{array}{c}
\vec{\theta}_{\uparrow} \\
\vec{\theta}_{\downarrow} 
\end{array}\right).
\end{equation}

Applying this to our action, we can complete this change of basis:

\begin{equation}
S_{CC} = \frac{1}{4\pi\beta}\sum_{\omega_{n}}|\omega_{n}|\left(\begin{array}{cc}
\vec{\theta}_{\uparrow} & \vec{\theta}_{\downarrow}
\end{array}\right)\left(\begin{array}{cc}
\mathbb{Y}\mathbb{g}_{c}\mathbb{Y} + \mathbb{I} & \mathbb{Y}\mathbb{g}_{c}\mathbb{Y} - \mathbb{I} \\
\mathbb{Y}\mathbb{g}_{c}\mathbb{Y} - \mathbb{I} & \mathbb{Y}\mathbb{g}_{c}\mathbb{Y} + \mathbb{I}
\end{array}\right)\left(\begin{array}{c}
\vec{\theta}_{\uparrow} \\
\vec{\theta}_{\downarrow}
\end{array}\right).
\end{equation}

The most general action, to first order in irrelevant electron tunneling processes, is then:

\begin{equation}
S = S_{CC} + \int_{0}^{\beta}\frac{d\tau}{\tau_{c}}V_{CC}.
\label{CCaction}
\end{equation}

To drive towards the $T_{0}=1,T_{\pi}=0$ fixed point, we need only take $v_{\pi}\rightarrow\infty$, as we have already established $v_{0}$ and $v_{\pi}$ as the quantities which push back along the $T_{0}$ and $T_{\pi}$ axes respectively.  

To massage something useful out of this strong-tunneling limit, we can utilize an extreme limit of the Villain approximation, following procedurally a longer formulation of the Kane-Fisher problem~\cite{kanefisher}.  Note that the partition function is the product of the contribution from $S_{CC}$ and a term of the form $e^{-v_{\pi}\int d\tau\cos\theta}$.  As $v_{\pi}\rightarrow\infty$, the entire partition function is zeroed out except when $\cos\theta$ is a minimum.  Acknowledging this, one can make the substitution of $e^{-v_{\pi}\int d\tau\cos\theta}\rightarrow\sum_{m}e^{im\theta}$ where $m$ is now a discrete step in time.  The partition function can then be integrated over $\theta$ once we complete the square to give a new Gaussian effective action in terms of the integer $m$.  Now, we can define $m=\partial_{t}\varphi/2\pi$, such that in the frequency domain, our bare action has \emph{the same form and an inverse Luttinger parameter}.  Hopping events between the minima of $\cos\theta$ are instantons whose term in the action takes the form $t\int d\tau\cos\varphi$, the same as if we had defined our electron operators as exponentials of $\varphi$ fields and examined tunneling about the bare action.   This equivalence between strong electron tunneling in $\theta$ and weak quasiparticle backscattering in $\varphi$ is a key feature of the Kane-Fisher problem for single-impurity scattering in Luttinger liquids.  Its physical implications, as well as a much more detailed derivation of it, can be found in Ref.~\onlinecite{kanefisher}.     

For four-terminal quantum point contacts, however, this equivalence manifests itself as a relationship between strong left-to-right electron tunneling and weak top-to-bottom electron backscattering~\cite{teokane}.  Therefore, our ultimate goal is to arrive at $\varphi_{\pi}$ operators which correspond to the $t_{\pi\rightarrow\pi}$ single-particle tunneling processes in Fig.~\ref{FigTV} about an M point in Fig.~\ref{FigRG}, for which only the $\pi$ band is pinched off.   

In our problem, the result of taking $v_{\pi}\rightarrow\infty$ and utilizing this trick is a fair bit more complicated due to the larger set of operators in the action and the matrix nature of the interactions. 

First, let us start by reducing this problem into manageable blocks.  We can simplify using the following definitions:

\begin{equation}
\mathbb{Y}\mathbb{g}_{c}\mathbb{Y}\pm\mathbb{I} = \mathbb{U}_{\pm}=\frac{1}{4}\left(\begin{array}{cc}
\mathbb{A}\pm 4\mathbb{I} & \mathbb{B} \\
\mathbb{B} & \mathbb{A}\pm 4\mathbb{I}
\end{array}\right) = \frac{1}{4}\left(\begin{array}{cc}
\mathbb{A}_{\pm} & \mathbb{B} \\
\mathbb{B} & \mathbb{A}_{\pm}
\end{array}\right)
\end{equation}

such that now

\begin{equation}
S_{CC} = \sum_{\omega_{n}}\frac{|\omega_{n}|}{4\pi\beta}\left[\sum_{\sigma=\uparrow,\downarrow}\vec{\theta}_{\sigma}^{T}\mathbb{U}_{+}\vec{\theta}_{\sigma} + 2\vec{\theta}_{\uparrow}^{T}\mathbb{U}_{-}\vec{\theta}_{\downarrow}\right]
\end{equation}

where

\begin{equation}
\mathbb{U}_{\pm}=\mathbb{Y}\mathbb{g}_{c}\mathbb{Y}\pm\mathbb{I}.
\end{equation}

As addressed in Eq.~(\ref{CCaction}), the full action contains contributions from $S_{CC}$ as well as perturbative tunneling terms.  To push towards the corner M phase, we can increase $v_{\pi}$ greatly until it becomes valid to replace that part of the partition function with discrete delta functions in $\vec{\theta}_{\pi\sigma}$:

\begin{equation}
\int_{0}^{\beta}\frac{d\tau}{\tau_{c}}V_{CC} = \frac{2i}{\beta}\sum_{\sigma=\uparrow,\downarrow}\vec{m}_{\pi\sigma}^{T}\vec{\theta}_{\pi\sigma} + \int_{0}^{\beta}\frac{d\tau}{\tau_{c}}\left[V_{0} + T_{\pi}\right]
\end{equation}

\begin{equation}
\vec{\theta}_{a\sigma}=\left(\begin{array}{c}
\theta_{a\sigma K} \\
\theta_{a\sigma K'}
\end{array}\right)
\end{equation}

such that $\vec{\theta}_{\sigma} = \vec{\theta}_{0\sigma}\oplus\vec{\theta}_{\pi\sigma}$.  $T_{\pi}$ is an instanton tunneling term which enforces the condition that $m_{a\sigma K/K'}$ is an integer.  The factor of $2$ in front of it is a normalization requirement from the $c,s\rightarrow\uparrow,\downarrow$ change of basis.  As described earlier, we can recognize the $m_{a\sigma K/K'}$ as discrete steps in time of a new field:

\begin{eqnarray}
\vec{m}_{\pi\sigma}&=&\left(\begin{array}{c}
m_{\pi\sigma K} \\
m_{\pi\sigma K'}
\end{array}\right) = \frac{1}{2\pi}\left(\begin{array}{c}
\partial_{t}\varphi_{\pi\sigma K} \\
\partial_{t}\varphi_{\pi\sigma K'} 
\end{array}\right) \nonumber \\
&\rightarrow& \frac{i\omega}{2\pi}\left(\begin{array}{c}
\varphi_{\pi\sigma K} \\
\varphi_{\pi\sigma K'}   
\end{array}\right) = \frac{i\omega}{2\pi}\vec{\varphi}_{\pi\sigma}.
\end{eqnarray}

Combining all of these definitions, we arrive at a full expression for the action which has both $\theta$ and $\varphi$ fields in it:

\begin{widetext}
\begin{eqnarray}
S &=&\bigg\{\sum_{\omega_{n}}\frac{|\omega_{n}|}{16\pi\beta}\bigg[\sum_{\sigma=\uparrow,\downarrow}\left(\sum_{a=0,\pi}\vec{\theta}^{T}_{a\sigma}\mathbb{A}_{+}\vec{\theta}_{a\sigma} + 2\vec{\theta}_{\pi\sigma}^{T}\mathbb{B}\vec{\theta}_{0\sigma}\right) + 2\sum_{a=0,\pi}\vec{\theta}_{a\uparrow}^{T}\mathbb{A}_{-}\vec{\theta}_{a\downarrow} + 2\vec{\theta}_{0\uparrow}^{T}\mathbb{B}\vec{\theta}_{\pi\downarrow} + 2\vec{\theta}_{0\downarrow}^{T}\mathbb{B}\vec{\theta}_{\pi\uparrow} \nonumber \\
&+& 16i\sgn(\omega_{n})\sum_{\sigma=\uparrow,\downarrow}\vec{\varphi}_{\pi\sigma}^{T}\vec{\theta}_{\pi\sigma}\bigg]\bigg\} + \int_{0}^{\beta}\frac{d\tau}{\tau_{c}}\left[V_{0} + T_{\pi}\right] 
\end{eqnarray}
\end{widetext}

where we have frequently exploited that $\mathbb{A}_{\pm}$ and $\mathbb{B}$ are symmetric.  

After a good bit of algebra, we can complete the square twice here and integrate out the $\vec{\theta}_{\pi\sigma}$, leaving us with an intermediate action $S_{M}$ about the $T_{0}=1$, $T_{\pi}=0$ M-phase stable fixed point:

\begin{equation}
S = S_{M} + \int_{0}^{\beta}\ \frac{d\tau}{\tau_{c}}V_{M}
\end{equation}

\begin{equation}
S_{M} = \frac{1}{4\pi\beta}\sum_{\omega_{n}}|\omega_{n}|\left(\begin{array}{cccc}
\vec{\theta}_{0\uparrow} & \vec{\theta}_{0\downarrow} & \vec{\varphi}_{\pi\uparrow} & \vec{\varphi}_{\pi\downarrow}
\end{array}\right)\mathbb{g}_{M}\left(\begin{array}{c}
\vec{\theta}_{0\uparrow} \\
\vec{\theta}_{0\downarrow} \\
\vec{\varphi}_{\pi\uparrow} \\
\vec{\varphi}_{\pi\uparrow} 
\end{array}\right)
\end{equation}

\begin{equation}
V_{M}=\sum_{\sigma,u}v_{0}\cos\left(\theta_{0\sigma u}\right) + t_{\pi}\cos\left(\varphi_{\pi\sigma u}\right)
\end{equation}

where $\sigma=\uparrow,\downarrow$; $u=K,K'$; and

\begin{widetext}
\begin{equation}
\mathbb{g}_{M}=\left(\begin{array}{cccc}
2\mathbb{I} + \alpha\sigma^{x} & \alpha\sigma^{x} & -\sgn(\omega_{n})\alpha\sigma^{x} & -\sgn(\omega_{n})\alpha\sigma^{x} \\
\alpha\sigma^{x} & 2\mathbb{I} + \alpha\sigma^{x} & -\sgn(\omega_{n})\alpha\sigma^{x} & -\sgn(\omega_{n})\alpha\sigma^{x} \\
\sgn(\omega_{n})\alpha\sigma^{x} & \sgn(\omega_{n})\alpha\sigma^{x} & 2\mathbb{I} - \alpha\sigma^{x} & -\alpha\sigma^{x} \\
\sgn(\omega_{n})\alpha\sigma^{x} & \sgn(\omega_{n})\alpha\sigma^{x} & -\alpha\sigma^{x} & 2\mathbb{I} - \alpha\sigma^{x}
\end{array}\right),\ \ \alpha=\frac{g-1}{g+1}
\end{equation}
\end{widetext}

in which the sign on the bottom right block of $\mathbb{g}_{M}$ is owed to one of the sets of bosonic fields being evaluated at $-\omega_{n}$ and the other at $+\omega_{n}$.  The tunneling term $t_{\pi}$ for the $\varphi$ fields represents weak, left-to-right tunneling of electrons with band index $\pi$; making it very large would return the system back to the CC corner fixed point.  

Correlation functions about this theory can be calculated using elements of the inverse of this matrix:

\begin{widetext}
\begin{equation}
2\mathbb{g}_{M}^{-1}=\frac{1}{2}\left(\begin{array}{cccc}
2\mathbb{I}  -\alpha\sigma^{x} & -\alpha\sigma^{x} & \sgn(\omega_{n})\alpha\sigma^{x} & \sgn(\omega_{n})\alpha\sigma^{x} \\
-\alpha\sigma^{x} & 2\mathbb{I} - \alpha\sigma^{x} & \sgn(\omega_{n})\alpha\sigma^{x} & \sgn(\omega_{n})\alpha\sigma^{x} \\
-\sgn(\omega_{n})\alpha\sigma^{x} & -\sgn(\omega_{n})\alpha\sigma^{x} & 2\mathbb{I} + \alpha\sigma^{x} & \alpha\sigma^{x} \\
-\sgn(\omega_{n})\alpha\sigma^{x} & -\sgn(\omega_{n})\alpha\sigma^{x} & \alpha\sigma^{x} & 2\mathbb{I} + \alpha\sigma^{x}
\end{array}\right)
\end{equation}
\end{widetext}

where the factor of $2$ again comes from the spin-index change of basis and ensures that diagonal correlators like $\ \langle e^{i\theta_{0\uparrow K}(\tau)}e^{i\theta_{0\uparrow K}(0)}\rangle\sim 1/\tau^{2}$ are appropriately marginal.  The calculations in the following section are also only appropriate provided that the single-particle, \emph{off-diagonal} correlation functions are zero.  This requirement in enforced for the $\theta$ fields for $2-\alpha>1$ and by $2+\alpha>1$ for the $\varphi$ variables.  However, the range $-1<\alpha<1$ is satisfied by all physical values of $g$.  Therefore, the only bounds on the validity of the $S_{M}$ theory are the regions in Fig.~\ref{FigPhase} for which many-body processes become relevant.

\subsection{Cubic-Order Fixed Points about the M Phase for $g_{-}\approx 1$} 
\label{appendix:cornerfixed}  

In this section, we will, using the $S_{M}$ theory established in the previous section, develop quartic-order flow equations for $v_{0}$ and $t_{\pi}$ for $g_{-}=1,\ g_{+}=g$.  After establishing the flow in those variables, we will relate $v_{0}$ and $t_{\pi}$ to $T_{0/\pi}$ in Eq.~(\ref{flowEQ}).  Expanding $g=1+\epsilon_{+}$, we will develop the cubic-order corrections to Fig.~\ref{FigQuad} close to the M-phase fixed points and obtain the schematic phase diagram Fig.~\ref{FigRG}.  

There are two sets of correlation functions for each coupling coefficient that we must calculate and rescale to obtain flow equations, as indicated by the two non-zero, off-diagonal elements for each operator and spin in $\mathbb{g}_{M}$.  For now, we will work just with the $v_{0}$ renormalization and then use band-index-exchange and pinch-off symmetries to relate them to the flow of $t_{\pi}$.  

Starting with the $\theta_{0\uparrow K}$ operator, we can calculate terms in the cumulant expansion to cubic order: 

\begin{widetext}
\begin{eqnarray}
\delta v_{0} &=& v_{0}^{3}\sum_{\sigma=\uparrow,\downarrow}\int_{-\infty}^{\infty}\ d\tau_{1}d\tau_{2}\langle T_{\tau}\left[e^{i\theta_{0\uparrow K}(\tau)}e^{i\theta_{0\sigma K'}(\tau_{1})}e^{-i\theta_{0\sigma K'}(\tau_{2})}\right]\rangle_{>} - \langle e^{i\theta_{0\uparrow K}(\tau)}\rangle_{>}\langle T_{\tau}\left[e^{i\theta_{0\sigma K'}(\tau_{1})}e^{-i\theta_{0\sigma K'}(\tau_{2})}\right]\rangle_{>} \nonumber \\
&+& v_{0}t_{\pi}^{2}\sum_{\sigma=\uparrow,\downarrow}\int_{-\infty}^{\infty}\ d\tau_{1}d\tau_{2}\langle T_{\tau}\left[e^{i\theta_{0\uparrow K}(\tau)}e^{i\varphi_{\pi\sigma K'}(\tau_{1})}e^{-i\varphi_{\pi\sigma K'}(\tau_{2})}\right]\rangle_{>} - \langle e^{i\theta_{0\uparrow K}(\tau)}\rangle_{>}\langle T_{\tau}\left[e^{i\varphi_{\pi\sigma K'}(\tau_{1})}e^{-i\varphi_{\pi\sigma K'}(\tau_{2})}\right]\rangle_{>}. \nonumber \\
\end{eqnarray}
\end{widetext}

As before, $T_{\tau}$ is the time-ordered product and for each of the integrals, the second term is the disconnected piece.  Taking into account all possible time orderings, we can rewrite this:

\begin{widetext}
\begin{eqnarray}
\delta v_{0} &=& 2v_{0}^{3}\int_{-\infty}^{\infty}\ d\tau_{1}d\tau_{2}\ \frac{1}{\tau_{12}^{2}}\left[\left(\frac{\tau_{1}}{\tau_{2}}\right)^{\alpha}-1\right] \nonumber \\
&+& 2v_{0}t_{\pi}^{2}\int_{-\infty}^{\infty}\ d\tau_{1}d\tau_{2}\ \frac{1}{\tau_{12}^{2}}\left[\Theta(\tau\not\in(\tau_{1},\tau_{2})) + e^{\pi i\alpha}\Theta(\tau_{1}<\tau<\tau_{2}) + e^{-\pi i\alpha}\Theta(\tau_{2}<\tau<\tau_{1}) -1\right]
\end{eqnarray}
\end{widetext}

where $\tau_{12}=\tau_{1}-\tau_{2}$ and $\Theta(\tau,\tau_{1},\tau_{2})$ is the Heaviside step function expressed in conditional notation.  These integrals can be performed analytically and, after a bit of work, we arrive at an explicit equation for $\delta v_{0}= v_{0}' - v_{0}$:

\begin{eqnarray}
\delta v_{0} = \log b\left[4v_{0}^{3}\pi\alpha\tan\left(\frac{\pi\alpha}{2}\right) - 8v_{0}t_{\pi}^{2}\sin^{2}\left(\frac{\pi\alpha}{2}\right)\right] \nonumber \\
\end{eqnarray}

where $b$ is the time-integral cutoff.  Rescaling $b\rightarrow be^{-l}$ and taking into account the flow equations' invariance under combined exchange $v_{0}\leftrightarrow t_{\pi},\ \alpha\leftrightarrow -\alpha$, we arrive at our higher-order flow equations:

\begin{eqnarray}
\frac{dv_{0}}{dl} &=& 4v_{0}^{3}\pi\alpha\tan\left(\frac{\pi\alpha}{2}\right) - 8v_{0}t_{\pi}^{2}\sin^{2}\left(\frac{\pi\alpha}{2}\right) \nonumber \\
\frac{dt_{\pi}}{dl} &=& 4t_{\pi}^{3}\pi\alpha\tan\left(\frac{\pi\alpha}{2}\right) - 8v_{0}^{2}t_{\pi}\sin^{2}\left(\frac{\pi\alpha}{2}\right)
\label{eqTV}
\end{eqnarray}

which are valid only in a perturbative vicinity of this M-phase fixed point and for weak-to-moderate interactions.

Now, to answer our initial question, we will examine the consequences of our new flow equations near the vicinity of the possible fixed line in Fig.~\ref{FigQuad}b.  Expanding $g=1+\epsilon_{+}$, $\alpha^{2}=\epsilon_{+}^{2}/4$ such that Eq.~(\ref{eqTV}) reduces to the $\epsilon_{-}=0$ limit of Eq.~(\ref{flowEQ}) under the substitution:

\begin{equation}
v_{0}=\frac{2}{\pi}\sqrt{1-T_{0}},\ t_{\pi}=\frac{2}{\pi}\sqrt{T_{\pi}}. 
\end{equation}

when $1-T_{0}\ll 1,\ T_{\pi}\ll 1$.  We also know that the linear terms in $dT_{0/\pi}/dl$ must agree with any local expansion in the $T_{0}-T_{\pi}$ plane, such that we can restore flow to linear order in $\epsilon_{-}$ near the M-phase by extracting the leading term in Eq.~(\ref{flowEQ}):

\begin{eqnarray}
\left. \frac{dT_{0}}{dl}\right|_{\mathcal{O}(\epsilon_{+}\epsilon_{-})} &=& 8\epsilon_{+}\epsilon_{-}(1-T_{0}) \nonumber \\
\left. \frac{dT_{\pi}}{dl}\right|_{\mathcal{O}(\epsilon_{+}\epsilon_{-})} &=& -8\epsilon_{+}\epsilon_{-}T_{\pi}
\end{eqnarray}

where we have used the same approximation $1-T_{0}\ll 1,\ T_{\pi}\ll 1$.  Combining this with an expansion of Eq.~(\ref{eqTV}) to quartic order in $\epsilon_{+}$, we obtain, finally, higher-order flow equations about an M-phase fixed point:

\begin{widetext}
\begin{eqnarray}
\frac{d}{dl}(1-T_{0}) &=&  -8\epsilon_{+}\epsilon_{-}(1-T_{0}) + 4(1-T_{0})^{2}\left[\epsilon_{+}^{2} + \frac{\pi^{2}\epsilon_{+}^{4}}{48}\right] - 4(1-T_{0})T_{\pi}\left[\epsilon_{+}^{2} - \frac{\pi^{2}\epsilon_{+}^{4}}{48}\right] \nonumber \\
\frac{dT_{\pi}}{dl} &=&  -8\epsilon_{+}\epsilon_{-}T_{\pi} + 4T_{\pi}^{2}\left[\epsilon_{+}^{2} + \frac{\pi^{2}\epsilon_{+}^{4}}{48}\right] - 4(1-T_{0})T_{\pi}\left[\epsilon_{+}^{2} - \frac{\pi^{2}\epsilon_{+}^{4}}{48}\right].
\end{eqnarray}
\label{quartFlow}
\end{widetext}

Equation~(\ref{quartFlow}) contains a few points of interest.  First and foremost, it reflects the pinch-off symmetry in its invariance under the exchange $(1-T_{0})\leftrightarrow T_{\pi}$.  It also contains fixed points at $T_{0}=1-\frac{2\epsilon_{-}}{\epsilon_{+}},\ T_{\pi}=0$ and at $T_{0}=0,\ T_{\pi}=\frac{2\epsilon_{-}}{\epsilon_{+}}$, in agreement with the small $\epsilon_{-}/\epsilon_{+}$ limit of Fig.~\ref{FigQuad}.  We specifically normalized our correlations about $S_{M}$ to fix this agreement, such that we could locate at quartic order in $\epsilon_{+}$ any local quantum critical points which control the transition between the CC/II and M phases.  

On the line $1-T_{0}=T_{\pi}$, Eq.~(\ref{quartFlow}) admits an additional fixed point at 

\begin{equation}
1-T_{0}=T_{\pi}=\frac{48}{\pi^{2}}\frac{\epsilon_{-}}{\epsilon_{+}^{3}}.
\end{equation}.

Relevant flow lines from this point head off towards the central $T_{0}=T_{\pi}=1/2$ critical point, \emph{even in the $\epsilon_{-}=0$ case when this point converges with the other critical points at the corner}.  Therefore, this point stands as a demonstration that the fixed line in Fig.~\ref{FigQuad}b is an artifact of $\mathcal{O}(\epsilon_{+}^{2})$ perturbation theory, at least in the vicinity of the $T_{0}=1,\ T_{\pi}=0$ M point.  It is most likely that this flow away from the M point continues, to lowest order, all the way to the central quantum critical point.  This infers that the phase diagram of this system for weak-to-moderate interactions is best described by the schematic Fig.~\ref{FigRG}.  


\end{appendix}

\end{document}